\documentclass[twocolumn,apj]{aastex63}
\usepackage{apjfonts}
\usepackage{float}
\usepackage[titletoc, title]{appendix}
\usepackage{amsmath}
\usepackage{textgreek}
\hypersetup{linkcolor=red,citecolor=blue,filecolor=red,urlcolor=blue}

\graphicspath{{./}{figures/}}

\shorttitle{\ion{Fe}{17} 2$p$-3$s$ Line Ratios}
\shortauthors{Grell, Leutenegger \& Shah 2021}

\begin{document}

\title{\large\textsc{\ion{Fe}{17} $2p\mbox{ -- }3s$ line ratio diagnostic of shock formation radius in O stars}}

\author[0000-0003-3363-9786]{Gabriel J. Grell}
\affiliation{University of Maryland College Park, College Park, MD 20742, USA}
\affiliation{NASA Goddard Space Flight Center, Greenbelt, MD 20771, USA}
\affiliation{CRESST II, Greenbelt, MD 20771, USA}

\author[0000-0002-3331-7595]{Maurice A. Leutenegger}
\affiliation{NASA Goddard Space Flight Center, Greenbelt, MD 20771, USA}

\author[0000-0002-6484-3803]{Chintan Shah}
\altaffiliation{NASA Postdoctoral Program Fellow}
\affiliation{NASA Goddard Space Flight Center, Greenbelt, MD 20771, USA}
\affiliation{Lawrence Livermore National Laboratory, Livermore, CA 94550, USA}
\affiliation{Max-Planck-Institut f\"ur Kernphysik, D-69117 Heidelberg, Germany}

\begin{abstract}

The $2p\mbox{ -- }3s$ lines of \ion{Fe}{17} in the X-ray spectrum of the O-type star $\zeta$~Puppis exhibit an anomalous (3G~+~M2)~/~(3F) line ratio of $\sim 1.4$, in comparison with $\sim 2.4$ for almost all other collisionally excited astrophysical spectra. Based on the work of \cite{MLF01}, we conjectured that the strong UV field of $\zeta$ Puppis produces the observed ratio by depopulation of metastable $3s$ excited states, and that the ratio can potentially be used as an independent diagnostic of plasma formation radius. We used the Flexible Atomic Code (FAC) collisional-radiative model to model the effect of UV photoexcitation from O stars on the \ion{Fe}{17} lines. We compared our model calculations to archival spectra of coronal and hot stars from the Chandra HETGS and XMM-Newton RGS to benchmark our calculations for various electron densities and UV field intensities. Our calculations show that UV photoexcitation does not produce a sufficiently large dynamic range in the 3F~/~(3F~+~3G~+~M2) fraction to explain the difference in the observed ratio between coronal stars and $\zeta$~Pup. Thus, this effect likely cannot explain the observed line ratio of $\zeta$~Pup, and its origin is still unexplained.

\end{abstract}


\section{Introduction}

The first discovery of X-ray emission from a massive hot star was achieved by the Einstein satellite in 1979 after observations of the X-ray binary Cyg X-3 revealed the presence of bright O-type stars in the nearby vicinity \citep{Hel79}. The X-ray spectra of hot stars are mainly thermal in nature, as they have been proven to be dominated by discrete lines from metals with ionization stages skewed toward lower temperatures \citep{GN2009}. The spectra also appear soft (particularly for O stars), as the best fits have favored thermal components with temperatures less than 1 keV \citep{ZP2007, Cohen2014}.

OB stars are known to produce powerful winds with mass-loss rates as high as $10^{-5}\, \mbox{M}_{\odot}\, \mbox{yr}^{-1}$ \citep{Morton67b, Pel06}. These winds are driven by radiation pressure from scattering in UV transitions \citep{CAK}, where this force is multiplied by displacement of optically thick driving transitions from their shadow in frequency space due to their Doppler shift in the supersonic wind, an effect known as deshadowing.

The currently accepted model for X-ray production in single, nonmagnetic OB stars was introduced by \cite{FPP97}, who used hydrodynamic simulations to show that X-ray emission arises in mutual collisions of dense, shock-compressed shells; this phenomenon is known as the embedded wind shock (EWS) mechanism. These wind shocks were theorized to arise from the instabilities intrinsic to deshadowing in the line-driving mechanism responsible for the stellar wind \citep{LS70, OCR88}.  The scattered radiation field should suppress these instabilities near the wind base, and shocks are expected to form starting a few tenths of the stellar radius above the photosphere \citep{OP1999, SundO2015}. 

 X-ray emission lines observed with the high-resolution diffraction grating spectrometers on board XMM-Newton and Chandra have confirmed the source of soft X-ray emission in single, nonmagnetic O stars to be EWS \citep{CMWMC01, Kel01, KCO03}, given their relatively soft spectra and velocity-broadened emission lines. \cite{OC01} calculated theoretical X-ray line profiles expected for hot stars when assuming that X-ray-emitting material follows the bulk motion of the wind, finding that higher continuum photoelectric absorption optical depths produce more asymmetric, blue-shifted lines. These models were applied to derive wind optical depths and thus make mass-loss rate estimates for a sample of stars observed with Chandra and XMM-Newton \citep{Cohen2010, Cohen2014}.

The forbidden-to-intercombination line ratio $\mathcal{R} \equiv {\it f/i}$ of helium-like ions is a diagnostic of electron density and UV field strength\footnote{In this article we adopt the notational convention that ratios are denoted with calligraphic $\mathcal{R}$, while stellar radii are denoted with italic $R$.}. UV photons and/or collisions depopulate the metastable upper level $1s 2s \, ^{3}S_{1}$ of the forbidden line ($1s 2s \, ^{3}S_{1} \rightarrow 1s^{2} \, ^{1}S_{0}$) and weaken it while enhancing the strength of the intercombination lines ($1s 2p \, ^{3}P_{1,2} \rightarrow 1s^{2} \, ^{1}S_{0}$) \citep{GJ1969, BDT72}. The scaling of the ratio with UV flux and electron density is given by

\begin{equation} 
\label{eqn:helike_ratio}
\mathcal{R} = {\mathcal{R}_0} \, \frac{1}{1 + \phi/\phi_{c} + n_{e}/n_{c}} \, ,
\end{equation}

\noindent where $\mathcal{R}_0$ is the ratio value in the limit of no UV photoexcitation and low electron density, $\phi$ is the photoexcitation rate from 2 $^{3}$S to 2 $^{3}$P, $n_{e}$ is the electron density, $\phi_{c}$ is the critical photoexcitation rate at which $\mathcal{R}$ = $\mathcal{R}_0$/2, and $n_{c}$ is the critical density. The mean density in the wind of a massive star is given by $n_e = 6.4 \times 10^9\, \mbox{cm}^{-3}\, \dot{M}_{-6}\, r^{-2}_{20}\, v^{-1}_{2000}$, where $\dot{M}_{-6}$ is the mass-loss rate in units of $10^{-6}\, \mbox{M}_\odot\, \mbox{yr}^{-1}$, $r_{20}$ is the radius in units of $20\, \mbox{R}_\odot$, and $v_{2000}$ is the velocity in units of $2000\, \mbox{km}\, \mbox{s}^{-1}$. Such densities are not high enough to affect $\mathcal{R}$ for most helium-like ions in most of the wind, although the density may be important for \ion{N}{6} and \ion{C}{5} in the inner wind. Because the mean UV field strength of massive stars is quite high, $\mathcal{R}$ can thus be used as a diagnostic for UV field strength and therefore shock location \citep{Kel01}.

\cite{LPKC06} incorporated the effect of the radial dependence of UV flux on $\mathcal{R} \equiv {\it f/i}$ in the context of line profiles based on \cite{OC01}, and used this to model the helium-like triplet ion complex to constrain the radial distribution of X-ray-emitting plasma in four OB stars. This was parameterized by a shock onset radius $R_0$, with X-ray-emitting plasma having a fixed filling factor above this radius. They found that the minimum onset radius of emission is typically 1.25 $<$ $R_0/R_{\star}$ $<$ 1.67, where $R_*$ is the stellar radius. Furthermore, they showed that the forbidden line is formed at large radii, while the intercombination line is enhanced where the UV flux is higher (close to the star), with the result that the forbidden line is comparatively broad while the intercombination line is somewhat more centrally peaked.

\cite{MLF01} developed a similar diagnostic using the metastable $3s$ levels of the important neon-like \ion{Fe}{17} ion. They applied their model to the cataclysmic variable (CV) EX Hydra, which is thought to have a high-density polar accretion flow; however, this model has not yet been applied as an observational diagnostic of UV field strength in astrophysics. OB stars are a natural candidate for such an application.

\begin{figure}[ht]
\epsscale{1.3}
\plotone{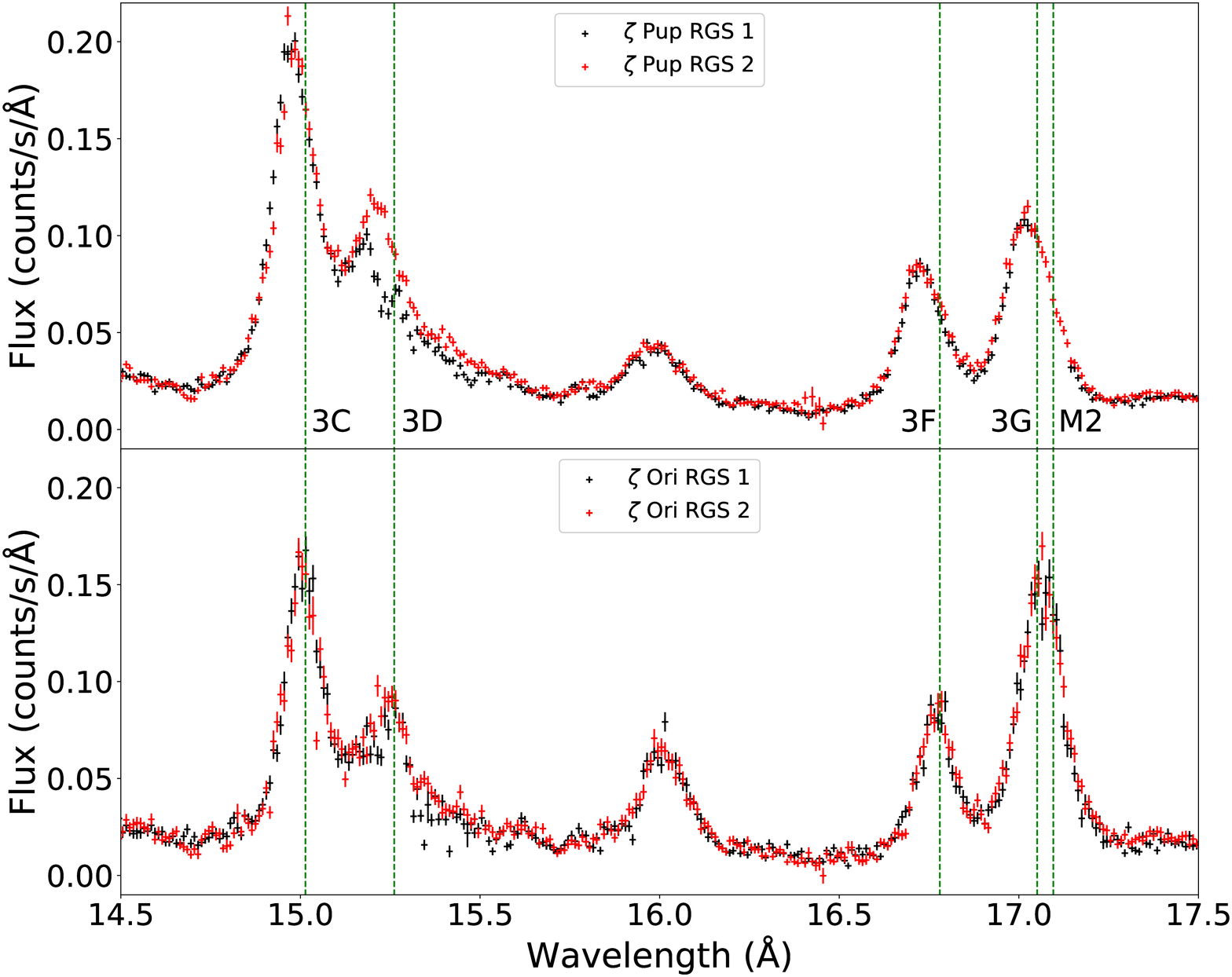}
\caption{XMM-Newton RGS spectra of $\zeta$~Pup (top) and $\zeta$~Ori (bottom). These spectra were obtained from the XMM-Newton Science Archive and reduced using the SAS (Science Analysis System) version 18.0. The black plus (+) signs represent the spectra from RGS1, and the red plus signs represent the spectra from RGS2. The green dashed lines represent the wavelengths of (left to right) the 3C, 3D, 3F, 3G, and M2 transitions. The M2+3G complex (17.06 - 17.14 $\mbox{\AA}$) is unresolved in both spectra due to Doppler broadening in the stars' supersonic winds.
\label{fig:rgs}}
\end{figure}

\cite{ML2012} noted an anomalous ratio in the \ion{Fe}{17} $2p-3s$ lines of the O4 supergiant $\zeta$~Puppis \citep[see also][]{2013A&A...551A..83H}. While the ratio (3G~+~M2)~/~3F is typically found to be approximately 2.4 for all O-type stars, as well as a range of other astrophysical sources, for $\zeta$~Pup it was found to be only about 1.4. \cite{Rauw2015} found a similar but lesser effect in the O6 supergiant $\lambda$~Cephei. Based on \cite{MLF01}, we conjecture that the strong UV fields of $\zeta$~Pup and $\lambda$~Cep could be responsible for the observed line ratios. However, because the M2+3G blend is typically unresolved in hot star spectra as shown in Figure~\ref{fig:rgs} for $\zeta$~Pup and $\zeta$~Ori, we cannot easily use the M2/3G ratio as in \cite{MLF01}, and we instead aim to develop the 3F~/~(3F~+~3G~+~M2) fraction as an independent diagnostic of plasma formation radius.

In this article, we use the Flexible Atomic Code (FAC) to calculate atomic data and a collisional-radiative model for \ion{Fe}{17}, accounting for the effect of UV photoexcitation from OB stars, and we incorporate the results of these calculations into existing line profile models for O-star X-ray spectra. \S~\ref{sec:fexvii} of this paper describes the metastable $3s$ levels of \ion{Fe}{17} and reviews the theoretical literature on this subject. \S~\ref{sec:atomic_model} describes the atomic model. \S~\ref{sec:lineprof} describes the line profile model {\tt newind}, which incorporates the radial dependence of the $2p-3s$ line ratios of neon-like ions. \S~\ref{sec:results} describes observations and data reduction for archival observations of coronal and hot stars taken by the Chandra HETGS and XMM-Newton RGS, and shows the model fitting results and calculations. In \S~\ref{sec:disc} we discuss our results and future work. 

\section{\ion{Fe}{17} diagnostics of UV field intensity and density}
\label{sec:fexvii}

\begin{figure}
\includegraphics[width=\columnwidth]{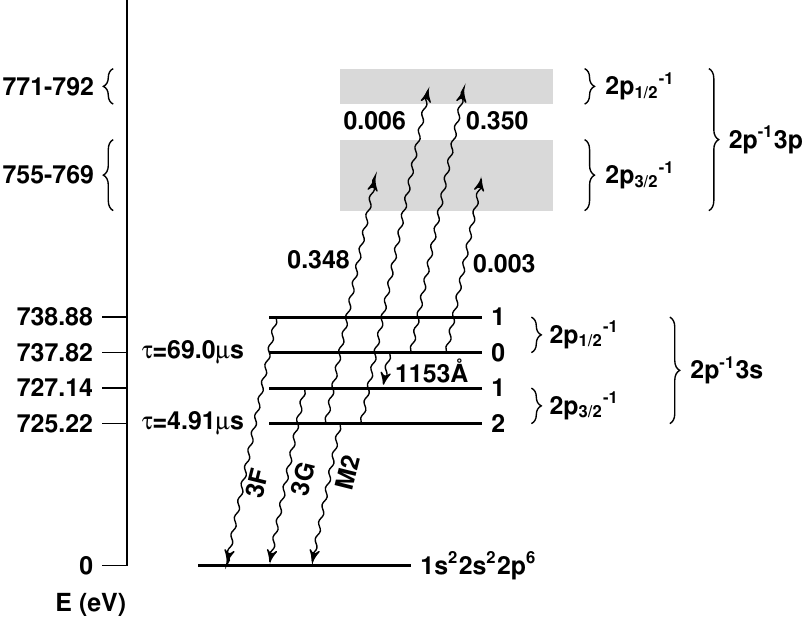}
\caption{Diagram showing important radiative processes for depopulating metastable $3s$ excited states of \ion{Fe}{17}. Individual energy levels are shown as horizontal lines, while for simplicity groups of $3p$ levels are shown as light-gray boxes. Wavy lines show radiative transitions. Radiative lifetimes of the two metastable states are shown \citep{CB2010, BCT2016}. Oscillator strengths are printed next to the corresponding transitions; these strengths represent sums over relevant levels.}
\label{fig:level_diagram}
\end{figure}

X-ray spectra from hot plasmas with temperatures of a few MK are dominated by the L-shell $3d\mbox{ -- }2p$ and $3s\mbox{ -- }2p$ transitions of \ion{Fe}{17} ions in the $15-17$ $\mbox{\AA}$ range \citep{par1973,chd2000,behar2001,xpb2002,PK2003}. The $3s\mbox{ -- }2p$ transitions known as the 3F, 3G, and M2 lines are produced by decay from [$2p_{1/2} 2p^{4}_{3/2} 3s_{1/2}$]$_{J=1}$ $^{3}P\,^{\circ}_{1}$, [$2p^{2}_{1/2} 2p^3_{3/2} 3s_{1/2}$]$_{J=1}$ $^{1}P\,^{\circ}_{1}$, and [$2p^{2}_{1/2} 2p^3_{3/2} 3s_{1/2}$]$_{J=2}$ $^{3}P\,^{\circ}_{2}$ to the [$2p^6$]$_{J=0}$ $^{1}S_{0}$ ground state, respectively, as depicted in Figure~\ref{fig:level_diagram}. The 3F line is observed at 16.777 $\mbox{\AA}$, the 3G line at 17.051 $\mbox{\AA}$, and the M2 line at 17.096 $\mbox{\AA}$ \citep{B98,May2005}.

\begin{table}
\centering
    \caption{\ion{Fe}{17} ground state and $2p^{-1} 3s$ and $2p^{-1} 3p$ singly excited levels and their respective configurations and energies (as calculated in this work using FAC and measured from \cite{B98} and \cite{BCT2016}). The 1$s$ and 2$s$ shells are closed in all configurations listed. LS coupling terms were assigned using the GRASP2K atomic structure package \citep{jo2013}. Mixing of LS coupling terms is especially strong between levels 2 and 4 ($J = 1$, odd), and levels 6, 9, and 13 ($J=2$, even). 
    \label{tab:lvls}}
    \begin{tabular}{ccccc}
    \hline\hline
      Level & \multicolumn{2}{c}{Configuration} & \multicolumn{2}{c}{Energy (eV)} \\ 
      & & & FAC & Experiment \\ 
      \hline
      0 & [$2p^6$]$_{J=0}$ & $^{1}S_{0}$ & 0.00 & 0.00\\
      1 & [$2p^{2}_{1/2} 2p^{3}_{3/2} 3s_{1/2}$]$_{J=2}$ & $^{3}P\,^{\circ}_{2}$ & 724.15 & 725.22\\
      2 & [$2p^{2}_{1/2} 2p^{3}_{3/2} 3s_{1/2}$]$_{J=1}$ & $^{1}P\,^{\circ}_{1}$ & 726.19 & 727.14\\
      3 & [$2p_{1/2} 2p^{4}_{3/2} 3s_{1/2}$]$_{J=0}$ & $^{3}P\,^{\circ}_{0}$ & 736.78 & 737.82\\
      4 & [$2p_{1/2} 2p^{4}_{3/2} 3s_{1/2}$]$_{J=1}$ & $^{3}P\,^{\circ}_{1}$ & 738.09 & 738.88\\
      5 & [$2p^{2}_{1/2} 2p^{3}_{3/2} 3p_{1/2}$]$_{J=1}$ &$^{3}S_{1}$ & 754.47 & \\
      6 & [$2p^{2}_{1/2} 2p^{3}_{3/2} 3p_{1/2}$]$_{J=2}$ &$^{3}D_{2}$ & 758.09 & \\
      7 & [$2p^{2}_{1/2} 2p^{3}_{3/2} 3p_{3/2}$]$_{J=3}$ &$^{3}D_{3}$ & 759.66 & \\
      8 & [$2p^{2}_{1/2} 2p^{3}_{3/2} 3p_{3/2}$]$_{J=1}$ &$^{1}P_{1}$ & 760.86 & \\
      9 & [$2p^{2}_{1/2} 2p^{3}_{3/2} 3p_{3/2}$]$_{J=2}$ &$^{3}P_{2}$ & 762.66 & \\
      10 & [$2p^{2}_{1/2} 2p^{3}_{3/2} 3p_{3/2}$]$_{J=0}$ &$^{3}P_{0}$ & 768.30 & \\
      11 & [$2p_{1/2} 2p^{4}_{3/2} 3p_{1/2}$]$_{J=1}$ &$^{3}D_{1}$ & 770.17 & \\
      12 & [$2p_{1/2} 2p^{4}_{3/2} 3p_{3/2}$]$_{J=1}$ &$^{3}P_{1}$ & 773.39 & \\
      13 & [$2p_{1/2} 2p^{4}_{3/2} 3p_{3/2}$]$_{J=2}$ &$^{1}D_{2}$ & 773.79 & \\
      14 & [$2p_{1/2} 2p^{4}_{3/2} 3p_{1/2}$]$_{J=0}$ &$^{1}S_{0}$ & 790.35 &\\
      \hline\hline
    \end{tabular}
\end{table}

\cite{MLF01} focused primarily on the strength of the M2 line as a density diagnostic, since its weakening is the most striking change in model spectra under the high-density conditions typical in the accretion flow of an intermediate polar cataclysmic variable CV. For most O stars, broadening of the emission lines makes it challenging to independently measure the strength of the M2 line, as it blends with 3G, although, because the broadening is typically only comparable to the splitting of 3G and M2, it is still possible. A cleaner diagnostic is possible using the (3G~+~M2)~/~3F ratio, although the dynamic range of this ratio across density and UV field strength is smaller than for the M2 line alone. For this reason we need to thoroughly optimize our calculations in two important ways. First, we need to use UV field strength models appropriate to O stars, rather than blackbody models. Second, we need to carefully consider the effects of systematic uncertainties in the atomic calculations themselves.

Numerous studies have demonstrated disparities between the astrophysical observations, laboratory experiments, and theoretical calculations of line intensity ratios of the $3s\mbox{ -- }2p$ transitions at the $\sim 10-20\%$ level \citep{Phil99, Bei2002, bbg2004}. 
Moreover, the theoretical predictions of the M2/3G line ratio in the low-density limit have been shown to yield significantly smaller values than both laboratory measurements and astrophysical observations of low-density plasmas~\citep{NS05, Gu2008}. 
Such discrepancies have sparked strong interest in the scientific community and given rise to a number of experimental and theoretical works attempting to explain the possible reason for the observed discrepancy.

Initially,~\citet{loulergue1975} and~\citet{smith1985} pointed out that the $3s$ line strengths can be affected by the resonant excitation (RE) in \ion{Fe}{17}. 
\citet{ssb1999} showed that dielectronic recombination (DR) in \ion{Fe}{18} can also alter the $3s$ level populations. 
Later, \cite{DB2002} utilized the Hebrew University Lawrence Livermore Atomic Code, a relativistic configuration interaction (RCI) code, to construct a three-ion model that included the effects of RE of \ion{Fe}{17}, DR of \ion{Fe}{18}, and collisional inner-shell ionization (CI) of \ion{Fe}{16} as line formation processes, leading to slightly better model-data agreement.
\citet{Gu2003} expanded this theoretical study to include all relevant L-shell ions (Fe~\textsc{xvii–xx}) using the relativistic distorted-wave method implemented in FAC and showed that DR and RE are highly important for Fe~\textsc{xvii–xx} in modeling of collisionally ionized plasmas. 
Both works also showed that 2$p$ inner-shell ionization affects the \ion{Fe}{17} $3s\mbox{ -- }2p$ transition by only 1-3 \%.

Furthermore, the collision strengths of $3s$ transitions were also investigated by~\citet{cpr2002} and~\citet{chen2003} using the Breit-Pauli {\it R}-matrix method. They included all possible resonance contributions that can arise from 89 atomic levels associated with the $n$ = 3 and 4 complexes in their close-coupling expansion.
\cite{Loch06} expanded their {\it R}-matrix calculations to 139 levels, including $2p^5 5l$ configurations. The generated atomic data were used in a collisional-radiative model to predict the $3s/3d$ line ratio. However, EBIT measurements~\citep{Bei2002} still showed $\sim20\%$ discrepancy with the {\it R}-matrix calculations. 
In order to better diagnose the theoretical origins of disagreement with experiment, rather than measuring line ratios as in~\citet{Bei2002,bbg2004}, \citet{bbc2006} measured line emission cross sections for \ion{Fe}{17} relative to the well-known radiative recombination cross sections and inferred that discrepancies between experiments, observations, and theories exist in the calculation of {\it direct} excitation cross sections. 
A converged Dirac {\it R}-matrix and relativistic distorted-wave calculations, reported by~\citet{che2007,chen2008}, with only $\sim5$\% error in calculated cross sections, showed $\sim20\%$ discrepancy with measurements~\citep{bbc2006}.

\cite{Gu2008} later reviewed the accuracy of previous \ion{Fe}{17} theories by comparing them to Chandra spectra of stellar coronae, finding that the main problem of the previous studies was likely the inability to fully include electron correlation effects, which are important for atomic structure calculations. 
In this work, second-order many-body perturbation theory (MBPT) was used to calculate highly accurate energy levels and transition matrix elements of \ion{Fe}{17} lines. The cross sections were calculated essentially using the distorted-wave method, though they have been corrected using the accurate multipole transition matrix elements that are calculated using the MBPT method. This improved the cross sections of \ion{Fe}{17} lines and reduced the disagreement with experiment from $\sim$20\% to $\sim$15\%. 

Recent studies have made progress in investigating these discrepancies. \cite{Gu2019} produced model spectra of ions from \ion{Fe}{17} to \ion{Fe}{24} for optically thin, collisionally ionized plasma. They expanded the work of~\citet{Gu2003} by including all relevant direct and indirect line formation processes. They used an updated version of FAC, which will be discussed in further detail in \S~\ref{sec:fac}. Their work yielded a 5$\%$ lower electron-impact RE rate for the M2 line and 30$\%$ lower resonant rate for the 3G line.

\cite{Shah19} used the Heidelberg FLASH-EBIT to produce an ion population mainly consisting of \ion{Fe}{17} ions in order to determine the $3s$ and $3d$ line emission cross sections. They improved the electron beam energy resolution ($\sim5$~eV) by an order of magnitude compared to previous experiments, allowing resolution of strong DR and RE resonance contributions to DE, and increased the counting statistics by three orders of magnitude compared to previous experiments~\citep{lkt2000,Bei2002,bbc2006,bro2008,glt2011}. 
Their cross section and line ratio measurements were compared to different combinations of theoretical atomic models. They found that the combination of distorted-wave and MBPT calculations, previously shown by~\citet{Gu2008}, led to good agreement with the total 3$s$ cross sections. The model also yielded a 9$\%$ discrepancy for the 3$d$ cross section and 11$\%$ for 3$s$/3$d$, which are both consistent with previous laboratory measurements~\citep{Bei2002,bbc2006}.
Moreover, these laboratory data were also used to calibrate the atomic data implemented in the SPEX~\citep{kaastra1996} spectral modeling code. Subsequently, these data were fed into a global model of the Chandra grating spectrum of Capella, which in turn improved the overall fit compared to the fit using the default data available in SPEX version 3.04~(see details in~\citet{Gu2019,gu2020}). 

A novel X-ray laser spectroscopy technique was also employed to directly scrutinize the underlying atomic structure of \ion{Fe}{17}. \citet{sbr2012} and~\citet{kuhn2020} measured the quantum mechanical oscillator strengths of \ion{Fe}{17} transitions and found them to be lower than predicted by most atomic theoretical calculations, but consistent with astrophysical observations and EBIT cross-section measurements. The theoretical calculations agreeing the best with the experiments were perturbation theory methods, as well as configuration interaction calculations using a very large number of states to achieve better convergence.

Despite these extensive efforts to improve theoretical methods and experiments over the past two decades, significant discrepancies still remain for \ion{Fe}{17} X-ray lines. Thus, as discussed in \S~\ref{sec:atomic_model}, our theoretical predictions for the (3G~+~M2)~/~3F ratio must therefore be interpreted in the context of aforementioned discrepancies.

\section{Atomic model}
\label{sec:atomic_model}
\subsection{FAC}
\label{sec:fac}

Our understanding of stellar astrophysics has made strides with the development of software packages capable of transforming how stellar theory and modeling interact with observations. One such software package is FAC, which calculates atomic structure, as well as a wide range of atomic radiative and collisional processes \citep{Gu2008}. The atomic code has proven to be robust, as excellent agreement has been found between FAC calculations and both astrophysical data and laboratory measurements \citep{Bitter2003, Gu2003, Zhong2004, FH2005, Gu2019}. 

FAC computes the atomic structure of the initial, intermediate, and final states of a particular charge state, along with its related transitions. The output data include energy levels, radiative and autoionization transition rates, collisional excitation and ionization cross sections, photoionization rates, and autoionization rates, thereby enabling the integration of various atomic processes within a single framework. The code implements a fully relativistic method based on the Dirac equation and distorted-wave approximation for continuum processes, which enable the ability to reliably model highly charged ions.

FAC is also equipped with a collisional-radiative model (CRM) that constructs synthetic spectra for plasmas under different physical conditions using atomic data. It calculates level populations for a given temperature and density assuming collisional-radiative equilibrium. These level populations are then multiplied by the radiative transition rates to derive line intensities.

We used FAC to calculate the line strengths of the \ion{Fe}{17} transitions at different UV field intensities in order to study changes in line ratio and model the effect of UV photoexcitation from O stars. To do this, we first used FAC to calculate the relevant atomic data for \ion{Fe}{17}, as well as for the neighboring charge states \ion{Fe}{18} and \ion{Fe}{16}, which contribute to the \ion{Fe}{17} line formation process through recombination and ionization. We included cascades up to $n=25$, and resonance excitation with spectators up to $n=10$. We then ran the CRM for a range of electron temperatures (see, e.g., Figure~\ref{fig:frac}) and a range of UV field configurations. For the model-data comparisons in \S~\ref{sec:results}, we assumed an electron temperature of 500 eV for all calculations. We assumed a density of $10^{4}$ cm$^{-3}$ for all calculations, i.e. in the limit of the low collisional excitation rate for metastable states.

FAC assumes a uniform UV field with mean local intensity $J_\nu$. For the case of an O star, it is typically assumed that the specific intensity $I_{\nu}$ is constant across the stellar disk (i.e., neglecting limb darkening, and neglecting the scattered radiation field in the wind), so that the mean intensity is given by
\begin{equation}
    J_{\nu}(r) = I_{\nu}(R_*)\, W(r)\, ,
    \label{eqn:Jnu}
\end{equation} 
where $W(r)$ is the geometrical dilution factor, i.e., the fractional solid angle subtended by the stellar disk from the point of view of a test particle at radius {\it r}: 
\begin{equation}
W(r) = \frac{1}{2} \left(1 - \left[1 - \left(\frac{R_{\star}}{r} \right)^{2}\right]^{1/2}\right)
\end{equation}
The maximum astrophysical dilution factor of $W(r) = 0.5$ represents the solid angle just above the photosphere; $W = 1$ would occur only in the interior of an ideal blackbody enclosure.

This is then used to calculate the photoexcitation rates: 
\begin{equation}
    \phi_{lu} = 4\pi \frac{\pi e^2}{m_e c} f_{lu} \frac{J_\nu}{h \nu}\, ,
    \label{eqn:phi}
\end{equation}
where $e$ is the electron charge, $m_e$ is the electron mass, $c$ is the speed of light, $f_{lu}$ is the oscillator strength for transitions from lower level $l$ to upper level $u$, $h$ is the Planck constant, and $\nu$ is the frequency of the transition.

\subsection{Analytical Model}
\label{sec:analytic}

In order to gain insight into the numerical results obtained with the FAC CRM and also to allow for adjustments to these results based on experimental and observational constraints, we sought to derive an analytical relation for the \ion{Fe}{17} line ratios as a function of UV field intensity. To do this, we solved the coupled rate equations for the $n=3$ singly excited states; a detailed discussion of this is given in Appendix~\ref{appendix:rate}. We obtained the following equations for the case where UV photoexcitation is important but the electron density is sufficiently low:

\begin{equation}
\mathcal{R}_1 = \frac{\mathcal{R}^\circ_1 (1 + P_3) + \mathcal{R}^\circ_3 P_{3,1}}{(1 + P_1) (1 + P_3) - P_{3,1} P_{1,3}}
\end{equation}
\begin{equation}
\mathcal{R}_2 = \mathcal{R}^\circ_2 + \mathcal{R}_1 P_{1,2} + \mathcal{R}_3 (1 + P_{3,2})\, ; 
\end{equation}
\begin{equation}
    \mathcal{R}_3 = \frac{\mathcal{R}^\circ_3 (1+ P_1) + \mathcal{R}^\circ_1 P_{1,3}}{(1 + P_1) (1 + P_3) - P_{3,1} P_{1,3}}\, ; 
\end{equation}
\begin{equation}
    \mathcal{R}_4 = \mathcal{R}^\circ_4 + \mathcal{R}_1 P_{1,4} + \mathcal{R}_3 P_{3,4} \, .
\end{equation}

Here the subscripts $i$ refer to the $3s$ excited states in ascending energy order from 1 to 4. $\mathcal{R}_i$ are the ratios of the strengths of decays from level $i$ to ground relative to the sum of all three lines for levels 1, 2, and 4, while $\mathcal{R}_3$ gives the strength of the UV transition from level 3 to level 2 relative to the sum of the three X-ray lines; in other words, $\mathcal{R}_1$, $\mathcal{R}_2$, and $\mathcal{R}_4$ give the fractional strengths of 3F, 3G, and M2, respectively, within the $2p-3s$ complex. $\mathcal{R}^\circ_i$ gives the line ratios in the absence of photoexcitation. $P_{i,j}$ gives the effective normalized photoexcitation rate from $3s$ level $i$ to $3s$ level $j$ summed over all intermediate $3p$ states, and $P_i $ gives the effective normalized photoexcitation rate from 3$s$ level {\it i} to all other levels combined. $P_{i,j}$ and $P_{i}$ are defined in \S~\ref{appendix:rate}.

Table~\ref{tab:lvls} shows the relevant \ion{Fe}{17} levels with their corresponding configurations and energies as calculated by us using FAC. Level index 0 is the ground state, 1-4 are the four 3$s$ excited states, and 5-14 are the 10 3$p$ excited states. 

Table~\ref{tab:transitions} shows the most important $2p \mbox{--} 3s$ transitions and their respective oscillator strengths, {\it $f_{lu}$}, and the branching ratios to the lower levels. These values were used to calculate the effective photoexcitation rates for the relevant levels. The final column is the product of the oscillator strength and branching ratio, showing the relative importance of each transition in changing the \ion{Fe}{17} level populations. 

\begin{figure}[ht]
\epsscale{1.3}
\plotone{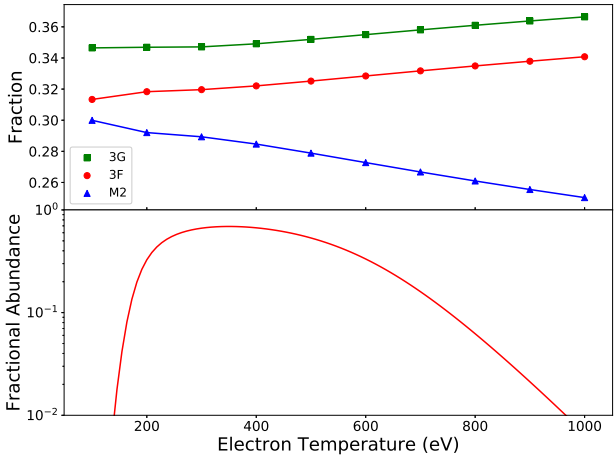}
\caption{Top: fractional line strengths as a function of electron temperature in the low-density limit ($n_e = 10^4$ cm$^{-3}$) and in the absence of UV photoexcitation for the 3F (red), 3G (green), and M2 (blue) transitions. The fractions vary only weakly with temperature. Bottom: neon-like ion fraction as a function of electron temperature. 
\label{fig:frac}}
\end{figure}

\begin{table}[ht!]
\centering
  \caption{Most important $3s\mbox{--}3p$ transitions and their respective oscillator strengths $f_{lu}$ and branching ratios $R$ to levels other than the original lower level. The product $f_{lu}\, R$ gives the effective strength for changing the $2p-3s$ line ratios. The transitions with the largest $f_{lu}\, R$, that are therefore the most relevant for changing observed line ratios, are shown in bold.
\label{tab:transitions}}
    \begin{tabular}{ccccc}
    \hline\hline
      Transition & $f_{lu}$ & Energy (eV) & $R$ & $f_{lu}\, R$\\ 
      \hline
      1$\rightarrow$5 & 5.22E-02 & 30.318 & 0.0417 & 2.17E-04\\
      3$\rightarrow$5 & 2.30E-03 & 17.893 & 0.997 & 2.30E-03\\
      {\bf 1$\rightarrow$6} & {\bf 5.35E-02} & {\bf 33.934} & {\bf 0.486} & {\bf 2.60E-02}\\
      1$\rightarrow$7 & 1.68E-01 & 35.505 & 7.96E-10 & 1.33E-10\\
      {\bf 1$\rightarrow$8} & {\bf 3.83E-03} & {\bf 36.705} & {\bf 0.939} & {\bf 3.60E-02}\\
      3$\rightarrow$8 & 7.06E-04 & 24.076 & 0.999 & 7.05E-04\\
      {\bf 1$\rightarrow$9} & {\bf 7.10E-02} & {\bf 38.510} & {\bf 0.424} & {\bf 3.01E-02}\\
      1$\rightarrow$11 & 7.53E-05 & 46.023 & 0.998 & 7.51E-05\\
      {\bf 3$\rightarrow$11} & {\bf 1.33E-01} & {\bf 33.394} & {\bf 0.572} & {\bf 7.61E-02}\\
      1$\rightarrow$12 & 5.16E-03 & 49.234 & 0.878 & 4.54E-03\\
      {\bf 3$\rightarrow$12} & {\bf 2.16E-01} & {\bf 36.605} &  {\bf 0.438} & {\bf 9.47E-02}\\
      1$\rightarrow$13 & 1.09E-03 & 49.639 & 0.983 & 1.07E-03\\
      \hline\hline
    \end{tabular}
\end{table}

\begin{figure}[ht]
\epsscale{1.3}
\plotone{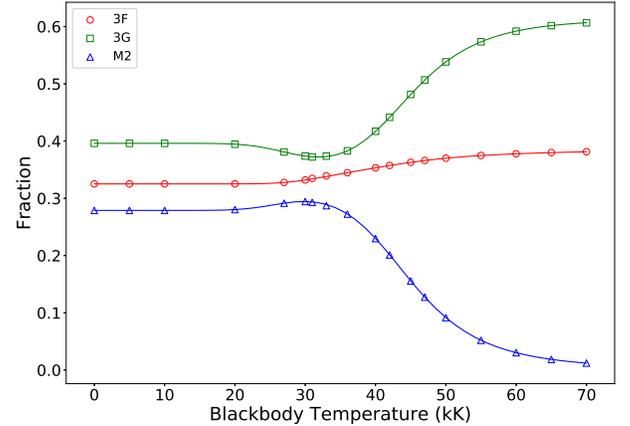}
\caption{Comparison of predicted \ion{Fe}{17} 3G (green), 3F (red), and M2 (blue) fractions as a function of blackbody temperature for $W=0.5$. The circles represent the FAC-CRM-predicted ratio values, and the lines represent the corresponding analytic model ratio values. 
\label{fig:crm}}
\end{figure}

\begin{figure}[ht]
\epsscale{1.3}
\plotone{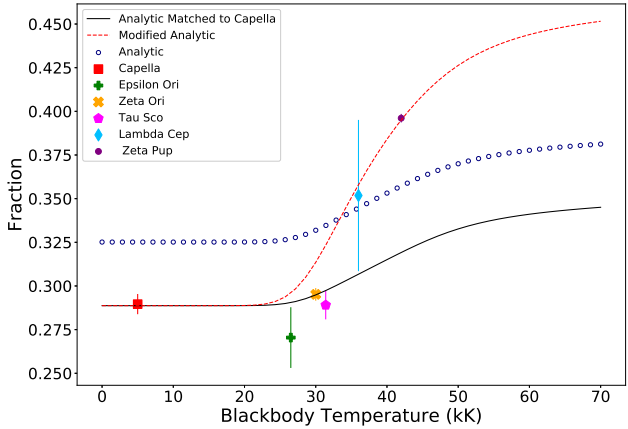}
\caption{Comparison of observed and calculated \ion{Fe}{17} 3F fractions as a function of photospheric blackbody temperature, evaluated at dilution $W = 0.5$, and with electron density $n_e = 10^4\, \mbox{cm}^{-3}$ and electron temperature $kT_e = 500$ eV. The analytical model fraction (red circles) did not match the observed fractions (points with error bars left to right) of Capella ($T_{\text{eff}}$ = 5 kK), $\epsilon$~Ori ($T_{\text{eff}}$ = 27 kK), $\zeta$~Ori ($T_{\text{eff}}$ = 30 kK), $\tau$~Sco ($T_{\text{eff}}$ = 31.4 kK), $\lambda$~Cep ($T_{\text{eff}}$ = 36 kK), and $\zeta$~Pup ($T_{\text{eff}}$ = 42.5 kK). The analytical model was adjusted so that the model ratio at low temperature matched that of Capella by changing the parameter $\mathcal{R}^0_4$ (black solid line); however, the dynamic range of the ratio was not sufficient to explain the observed ratio of $\zeta$~Pup. We further adjusted the model to match the observed ratio of $\zeta$~Pup by increasing $\mathcal{R}^0_3$ by a factor of 4.125 at the expense of $\mathcal{R}^0_2$ (red dashed line). This degree of increase required in $\mathcal{R}^0_3$ is unrealistic.
\label{fig:crm_3F}}
\end{figure}

\begin{figure}[ht]
\epsscale{1.3}
\plotone{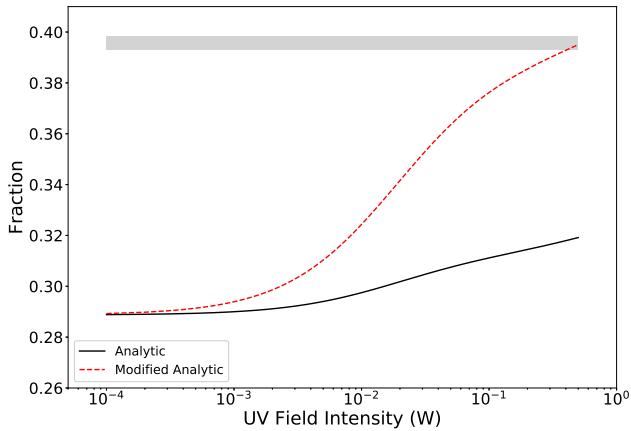}
\caption{\ion{Fe}{17} 3F fractions as a function of UV field intensity (W) for a 42,500 K blackbody (black solid line: analytic; red dashed line: analytic with ad hoc adjustment to $\mathcal{R}^0_3$). The gray stripe represents the envelope of the 3F fraction observed in $\zeta$ Pup.
\label{fig:FACvsW}}
\end{figure}

Figure~\ref{fig:frac} shows the trends of the fractions and neon-like charge balance in the absence of UV photoexcitation as a function of electron temperature (eV). The fractions vary only weakly as a function of this potentially confounding variable. 

Figure~\ref{fig:crm} compares the FAC CRM model and analytical model for the \ion{Fe}{17} 3G, M2, and 3F fractions as a function of temperature for a blackbody radiation field. The values of $\mathcal{R}_i^0$ in the analytical model were set from the FAC CRM calculations, but the dependence on UV flux was calculated using the FAC atomic data and blackbody flux. As expected, both plots exhibit an increase in the 3G line strength and a decrease in M2 line strength as temperature increases. In physical terms, UV photons are depopulating the upper M2 levels and therefore weakening it, while simultaneously enhancing the upper 3F and 3G levels. The small decrease in 3G around 30 kK is due to depopulation of level 3, the $J=0$ metastable state, which has a lifetime about an order of magnitude longer than level 1. The dependence of the fractions from the analytical models on blackbody temperature is in excellent agreement with the FAC-calculated fractions. 

Figure~\ref{fig:crm_3F} shows a comparison of the same FAC CRM and analytical model 3F fractions as a function of blackbody temperature; the analytical model fractions were adjusted in two ways. First, the value of $\mathcal{R}_4^0$ was decreased to match the observed fraction of Capella, consistent with the findings of previous studies \citep{Loch06, Gu2008}; then, we made a large ad hoc increase in the values of $\mathcal{R}_3^0$, such that the dynamic range of the 3F fraction model would match that observed between Capella and $\zeta$~Pup. The factor of 4.125 increase in $\mathcal{R}_3^0$ required to produce such an effect is very unrealistic and is included to illustrate the point that theoretical uncertainties in the level populations at low UV flux are likely not sufficient to explain the observations.

Figure~\ref{fig:FACvsW} compares the same analytic model curves as in Figure~\ref{fig:crm_3F} to the observed ratio of $\zeta$~Pup, but as a function of geometrical dilution $W(r)$ for a blackbody with $T_{\text{eff}} = 42.5$ kK. The \ion{Fe}{17} line emission from $\zeta$~Pup likely occurs throughout the wind over a range of dilution factors, but with the strongest weighting for the relatively large dilution factors occurring within a few stellar radii, where the wind density is largest \citep{LPKC06}.

\subsection{Model Atmospheres}

In \S~\ref{sec:analytic}, we used the FAC CRM to calculate the strengths of the 3G, 3F, and M2 lines of \ion{Fe}{17} for different values of geometrical dilution $W$ for blackbodies with a range of temperature, and we derived fractional strengths for each line by dividing by the sum of all three line strengths (i.e., 3F~/~[3F~+~3G~+~M2]). 
Real stellar spectra are much more complex than the blackbodies typically used to model the effects of UV radiation fields on these ions in studies of atomic physics. We thus sought to further optimize our calculation by using UV field strength models more appropriate to OB stars. 

Because most OB stars lie on sight lines having significant neutral interstellar gas, their EUV fluxes are not directly observable, and we must rely on model atmospheres. To this end, we utilized the TLUSTY OSTAR2002 and BSTAR2006 model atmosphere databases \citep{LH03, LH2007}. The OSTAR2002 grid contains datasets of spectral energy distributions (SEDs) typical of O stars covering a temperature range of 27.5 - 55 kK for the full set of frequency points used to calculate the model atmospheres. Similarly, the BSTAR2006 grid contains datasets of SEDs for B stars covering a 15 - 30 kK temperature range. UV fluxes from TLUSTY model atmospheres with effective temperatures ranging from 20 to 55 kK were used for the photoexcitation modeling. For $T_{\text{eff}}$ < 27 kK, we used BSTAR2006 datasets with a surface gravity of log g = 3.00, and for $T_{\text{eff}}$ > 27 kK we used OSTAR2002 datasets with a surface gravity $\log g$ = 4.00.


\begin{figure*}[ht]
\includegraphics[scale=0.6]{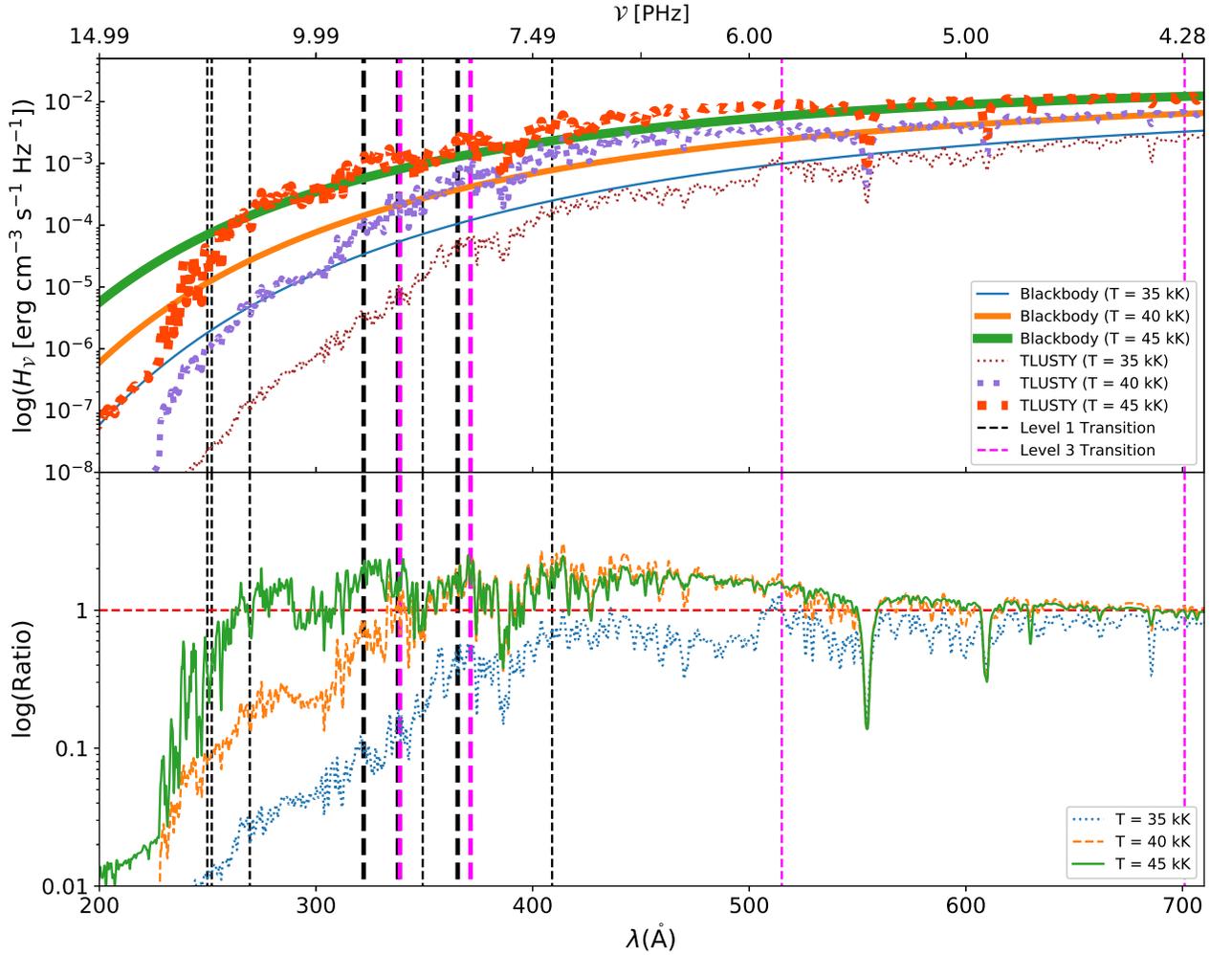}
\caption{Top: comparison of {\it T} = 35, 40, 45 kK blackbody models (blue, orange, green respectively) and {\it T} = 35, 40, 45 kK, log g = 3.50, Gaussian-broadened ({\it v} = 10 km s$^{-1}$) stellar atmosphere models from TLUSTY (red, purple, brown, respectively) as a function of wavelength ($\mbox{\AA}$). Bottom: ratio of TLUSTY models to blackbody models with the same effective temperature as a function of wavelength. The vertical lines represent the wavelength of the \ion{Fe}{17} $3s\mbox{--}3p$ transitions, with those originating from level 1 in black and those from level 3 in pink. The most important transitions are in boldface. The top axis shows the corresponding frequency range.
\label{fig:fac}}
\end{figure*}

\begin{figure}[ht]
\epsscale{1.3}
\plotone{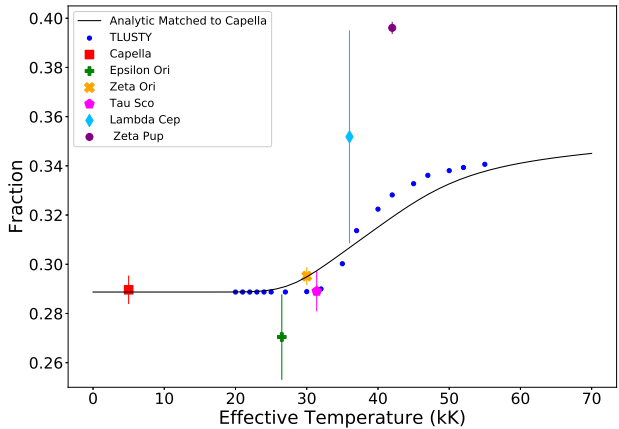}
\caption{Comparison of observed and calculated \ion{Fe}{17} 3F fractions as a function of stellar effective temperature. The black solid line represents the analytical model fraction from a blackbody, as previously shown in Figure~\ref{fig:crm_3F}, while the blue points represent the fractions calculated using TLUSTY model atmospheres. The TLUSTY fractions match the trends predicted by the UV flux ratios shown in Figure \ref{fig:fac}, as they are less than the analytical model ratios for {\it T} $<$ 35 kK and are greater for {\it T} $>$ 40 kK.
\label{fig:tlusty_model}}
\end{figure}

The top panel of Figure~\ref{fig:fac} illustrates the comparison between the blackbodies at various effective temperatures as a function of wavelength ($\mbox{\AA}$), while the bottom panel shows the ratio of the TLUSTY models to blackbody models for each effective temperature. The important $3s \mbox{--} 3p$ transition wavelengths calculated with FAC are shown on the figure as vertical lines, with the most important subset shown in bold face. There are significant differences between the TLUSTY models and blackbody models of the same temperature, particularly in the 200 - 228 $\mbox{\AA}$ range (shortward of the \ion{He}{2} ionization edge). However, the relevant $3s \mbox{--} 3p$ transition wavelengths are all longward of the \ion{He}{2} edge. For 45 kK, the TLUSTY models have up to a factor of two more flux than the corresponding blackbody at the relevant wavelengths, while for 35 kK, the fluxes range from comparable to more than an order of magnitude less than the blackbody. Because of these differences, we compared both FAC CRM calculations and our analytical model using both blackbodies and TLUSTY model atmospheres, as shown in Figure~\ref{fig:tlusty_model}. While the 3F fraction does show a somewhat steeper temperature dependence for TLUSTY models than for blackbodies, the basic behavior of the ratio as a function of temperature is unchanged.

\section{Line profile model}
\label{sec:lineprof}

The X-ray emission-line Doppler profiles of O stars have been successfully modeled by \cite{OC01}, with applications to Chandra and XMM-Newton spectra in, e.g.,  \cite{Cohen2010} and \cite{Cohen2014}. The key parameter in this model is the characteristic optical depth $\tau_* \equiv \kappa \dot{M} / 4 \pi v_\infty R_*$; here $\kappa$ is the opacity of the dominant unshocked part of the wind, mainly due to continuum photoelectric absorption in few times ionized metals, $\dot{M}$ is the mass-loss rate, $v_\infty$ is the wind terminal velocity, and $R_*$ is the stellar radius.

\cite{LPKC06} extended this model to the K$\alpha$ transitions of helium-like ions by incorporating the radial dependence of the forbidden-to-intercombination line ratio as affected by UV photoexcitation, as in Equation~\ref{eqn:helike_ratio}. 

We implemented a similar model to calculate line profiles for neon-like ions while including the radial dependence of the 3F, 3G, and M2 fractions. As in the case of the {\tt windprof} and {\tt hewind} models based on \cite{OC01} and 
\cite{LPKC06}, the new {\tt newind} model is implemented as an additive XSPEC local model. 

{\tt newind} can be used in one of two modes. In the first mode, the fractions are computed using a lookup table calculated directly with the FAC CRM module. In the second mode, the fractions are computed analytically, as in \S~\ref{sec:analytic}.

\begin{figure}[ht]
\epsscale{1.3}
\plotone{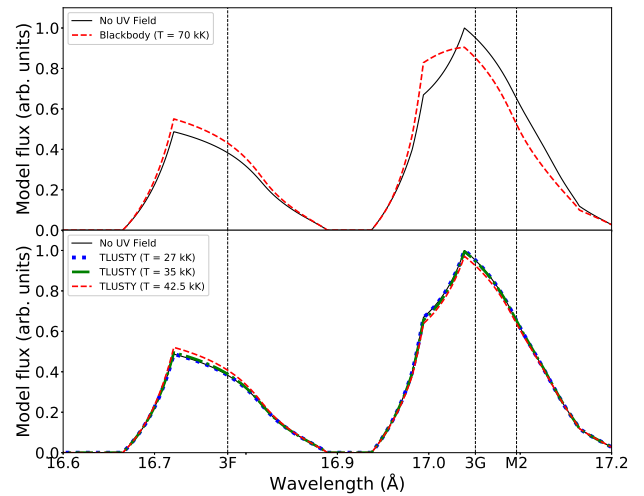}
\caption{Comparisons of {\tt newind} models with differing effective temperatures. Top: comparison of line profile with no UV field versus a blackbody with an effective temperature of 70 kK. Bottom: comparison of line profile with no UV field versus line profiles of TLUSTY model atmospheres with typical O-star effective temperatures (27.5, 35, 42.5 kK).
\label{fig:newind}}
\end{figure}

Figure~\ref{fig:newind} shows comparisons of {\tt newind} models with the same nominal parameters but different effective temperatures. The top panel compares a line profile with no UV field to a 70 kK blackbody calculated using the {\tt newind} analytic model mode. We used the following fiducial parameters: $\tau_{*}$ = 1, which is the characteristic continuum optical depth of the wind as defined in \cite{OC01}; X-ray emission onset radius $R_0$ = 2$R_{*}$; and wind terminal velocity $v_{\infty}$ = 2000 km s$^{-1}$. The bottom panel compares the same line profile with no UV field to TLUSTY model atmospheres with typical O-star effective temperatures: 27.5, 35, and 42.5 kK. For each TLUSTY model, we used $\log g = 4.00$ . 

As can be seen in the top panel, there is a noticeable but modest effect on the line profiles for the 70 kK blackbody, particularly in the M2 + 3G blend. However, the UV field from typical O stars has only a weak effect on the line profiles and ratios, as shown in the bottom panel.

\section{Model Fitting and Results}
\label{sec:results}
\subsection{Observations and Data Reduction}


We compared our model calculations to archival observations of coronal and hot stars taken by the Chandra HETGS and XMM-Newton RGS. We obtained archival X-ray spectra of the stars Capella, $\tau$~Scorpii, and EX~Hydra as they provide well-resolved spectral lines and are statistically robust. We also obtained spectra of the O stars $\zeta$~Orionis, $\epsilon$~Orionis, $\zeta$~Puppis, and $\lambda$~Cephei for the purpose of investigating the anomalous \ion{Fe}{17} $2p\mbox{ -- }3s$ line ratios in OB supergiants. 

HETGS spectra of Capella, $\tau$~Sco, EX~Hya, $\zeta$~Ori, and $\epsilon$~Ori were obtained from the Chandra archive and reprocessed using CIAO (Chandra Interactive Analysis of Observations) version 4.11 and CALDB (calibration database) version 4.8.4.1. RGS spectra of $\zeta$~Ori, $\tau$~Sco, $\epsilon$~Ori, $\lambda$~Cep and $\zeta$~Pup were obtained from the XMM-Newton Science Archive and reduced using SAS (Science Analysis System) version 18.0. Table~\ref{tab:starprops} shows a log of every star with their respective key physical parameters. The full list of spectral OBsIDs with corresponding exposure times is shown in Table~\ref{tab:obsid_appendix} in Appendix~\ref{appendix:obsid}. The spectra were fit using XSPEC version 12.10.1f \citep{Dorman2003} using the {\tt migrad} minimizer and the {\tt cstat} fit statistic \citep{C79}, which is appropriate for data following Poisson statistics. 

\begin{table}[ht!]
\centering
  \caption{Log of key stellar parameters. The list of OBsIDs with corresponding exposure times is shown in Appendix~\ref{appendix:obsid}. \string^ - UV flux of EX Hya may be higher than indicated by effective temperature \citep{MLF01}. Effective temperatures and surface gravities are taken from sources cited in the last column: N2003: \cite{Ness03}; P2016: \cite{Puebla2016}; R2008: \cite{Raassen2008}; M2001: \cite{MLF01}, D2006: \cite{Del06}; R2015: \cite{Rauw2015}; L1993: \cite{LL93}.
\label{tab:starprops}}
    \begin{tabular}{cccc}
    \hline\hline
      Star & $T_{\text{eff}}$ & log {\it g} & Ref.\\ 
       & (kK) & (cm s$^{-2}$)\\
      \hline
      Capella & 5 & - & N2003\\
      $\epsilon$ Orionis & 27 & 3.00 & P2016\\
      $\zeta$ Orionis & 29.5 & 3.25 & R2008\\
      EX Hydrae & 30\string^ & - & M2001\\
      $\tau$ Scorpii & 31.4 & 4.24 & D2006\\
      $\lambda$ Cephei & 36 & 3.50& R2015\\
      $\zeta$ Puppis & 42.5 & 3.75 & L1993\\
      \hline\hline
    \end{tabular}
\end{table}

We fit the archival spectra with a series of Gaussian models in order to derive values for the line strengths of the \ion{Fe}{17} 3F (16.777 $\mbox{\AA}$), 3G (17.051 $\mbox{\AA}$), and M2 (17.096 $\mbox{\AA}$) $2p\mbox{ -- }3s$ transitions \citep{B98, May2005}. We used 16.777 $\mbox{\AA}$ as the rest wavelength for the 3F transition, rather than the value of 16.780 $\mbox{\AA}$ given in \cite{B98}, as it agrees better with the observed wavelength of the transition in Capella.

The fits of Capella, $\tau$~Sco, and EX~Hya act as a benchmark of the FAC calculations; Capella benchmarks FAC in the limit of low density and UV flux; $\tau$~Sco provides a check of the line ratios at modest UV flux; and EX~Hya benchmarks the high-density regime (although the UV flux may also be nonnegligible). The four OB supergiants test the 3F line strength as a function of photospheric UV flux.

\S~\ref{subsec:cap} describes the fitting results for Capella, $\tau$~Sco, and EX~Hya, and \S~\ref{subsec:zeta} describes the results for $\zeta$~Ori, $\epsilon$~Ori, $\zeta$~Pup, and $\lambda$~Cep.

\subsection{Fitting Results}
\subsubsection{Capella, \texorpdfstring{$\tau$}{Tau}~Sco, and EX~Hya}
\label{subsec:cap}

\begin{figure}[ht]
\epsscale{1.3}
\plotone{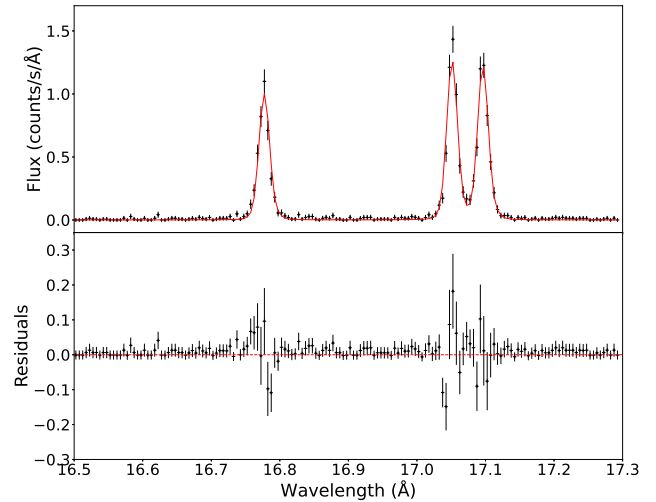}
\caption{Capella MEG spectrum (black) fit with Gaussian models (red). The line strengths at 16.777, 17.051, and 17.096 $\mbox{\AA}$ were used to derive the \ion{Fe}{17} $2p\mbox{ -- }3s$ line ratios. There is no UV photoexcitation effect on the line strengths of Capella, making it a robust benchmark of this limit.
\label{fig:capella}}
\end{figure}

Figure~\ref{fig:capella} shows the Gaussian fits to X-ray spectra of the coronal star Capella from Chandra observations. 
Capella is a coronal star ($T_{\text{eff}}$ = 5000 K) that has low UV flux and sufficiently low densities to serve as a benchmark for \ion{Fe}{17} line ratios. 
We fit the spectra of three different Chandra observations of Capella in order to estimate the line strengths of the 3F, 3G, and M2 transitions. The results are mutually consistent, and the ratio we used in this work is a weighted average of these results.

$\tau$~Sco is a massive magnetic \citep{Del06} B0V star ($T_{\text{eff}}$ = 31400 K) with well-resolved spectral lines \citep{Cel03, Mel03}, making it a robust benchmark source. Furthermore, because of its nonnegligible UV flux, it is a good candidate to search for a possibly heretofore-overlooked weak effect on the line ratios. However, upon comparison, the star yielded very similar line ratios to Capella. 

We also obtained line strengths from the spectra of the intermediate polar EX~Hydra to benchmark our fraction calculations in the limit of high electron density ($n_e \geq 3 \times 10^{14}$ cm$^{-3}$) and high UV field intensity. There was indeed good agreement between the FAC-predicted fraction calculations and the observed EX~Hya 3F, 3G, and M2 fractions at high density, thus reaffirming the results of \cite{MLF01}.

\subsubsection{\texorpdfstring{$\zeta$}{Zeta} Pup, \texorpdfstring{$\lambda$}{Lambda} Cep, \texorpdfstring{$\zeta$}{Zeta} Ori, and \texorpdfstring{$\epsilon$}{Epsilon} Ori}
\label{subsec:zeta}

\begin{figure}[ht]
\epsscale{1.3}
\plotone{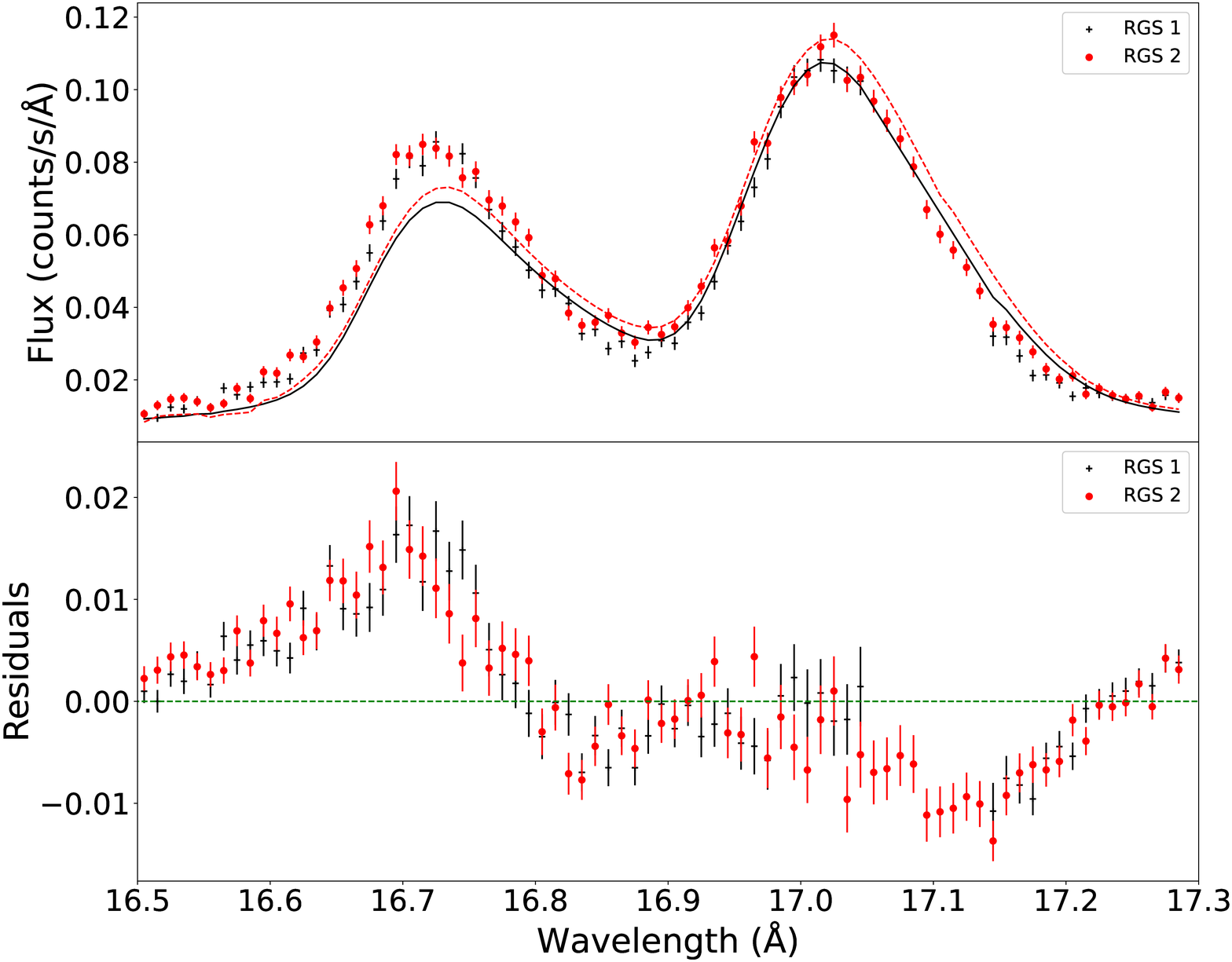}
\caption{$\zeta$ Pup spectra from the XMM-Newton RGS fit with the {\tt newind} model in analytical mode. The black plus signs represent the spectra from RGS1, and the red circles represent the spectra from RGS2.
\label{fig:zetapup}}
\end{figure}

Figure~\ref{fig:zetapup} shows our fits to the XMM-Newton RGS spectra of $\zeta$~Pup using {\tt newind}. For this, we used the closest TLUSTY model to the estimated stellar parameters of $\zeta$~Pup, with $T_{\text{eff}} = 42.5$ kK and $\log g = 4.00$. We also used values for $\mathcal{R}_i^\circ$ fixed to the values of Capella. As expected, the model underpredicted the flux of 3F while overpredicting the blend of 3G and M2. 

\begin{figure}[ht]
\epsscale{1.3}
\plotone{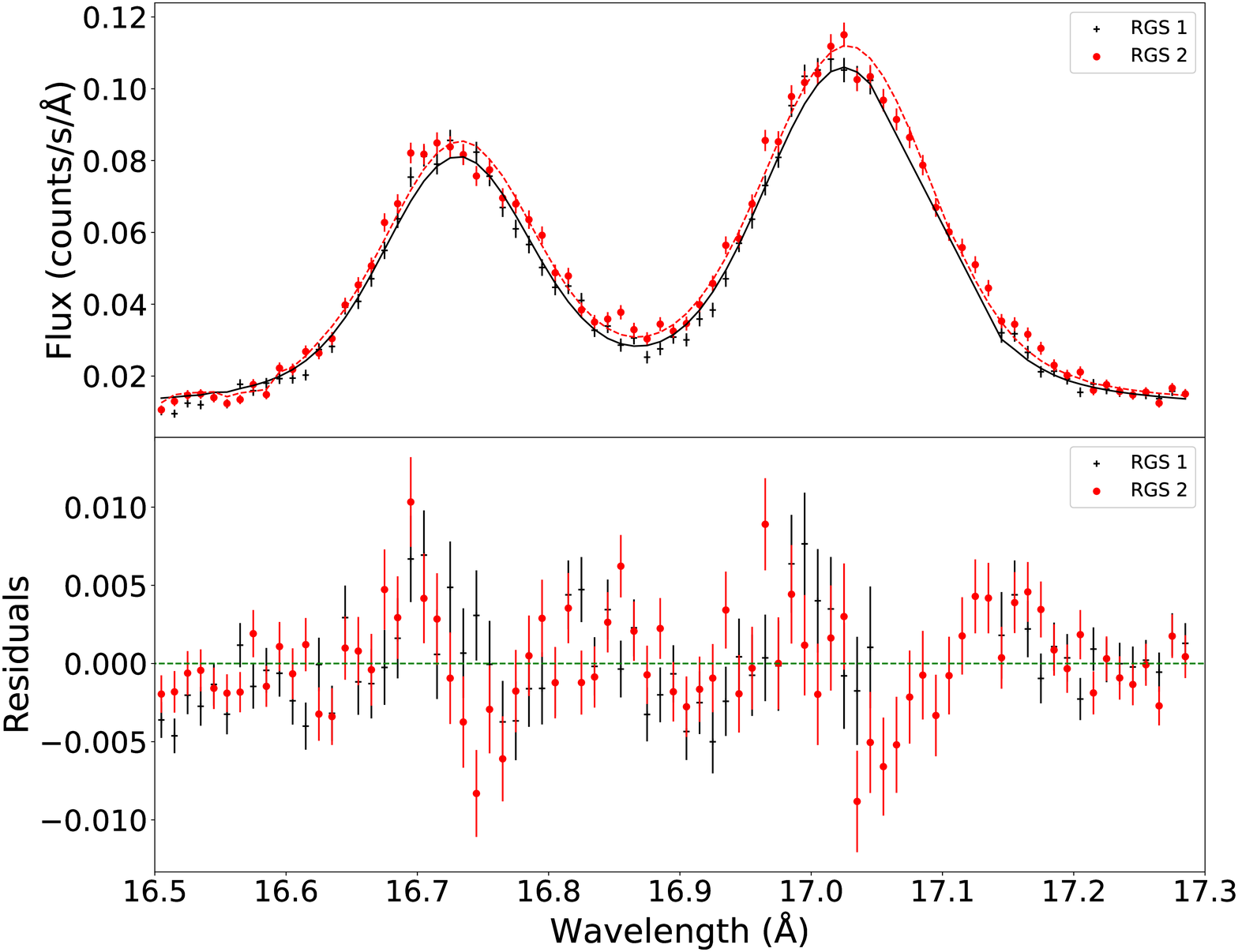}
\caption{$\zeta$ Pup spectra from the XMM-Newton RGS fit with Gaussian models. The black plus signs represent the spectra from RGS1, and the red circles represent the spectra from RGS2.
\label{fig:zpgauss}}
\end{figure}

\begin{figure*}[ht]
\epsscale{1.25}
\plotone{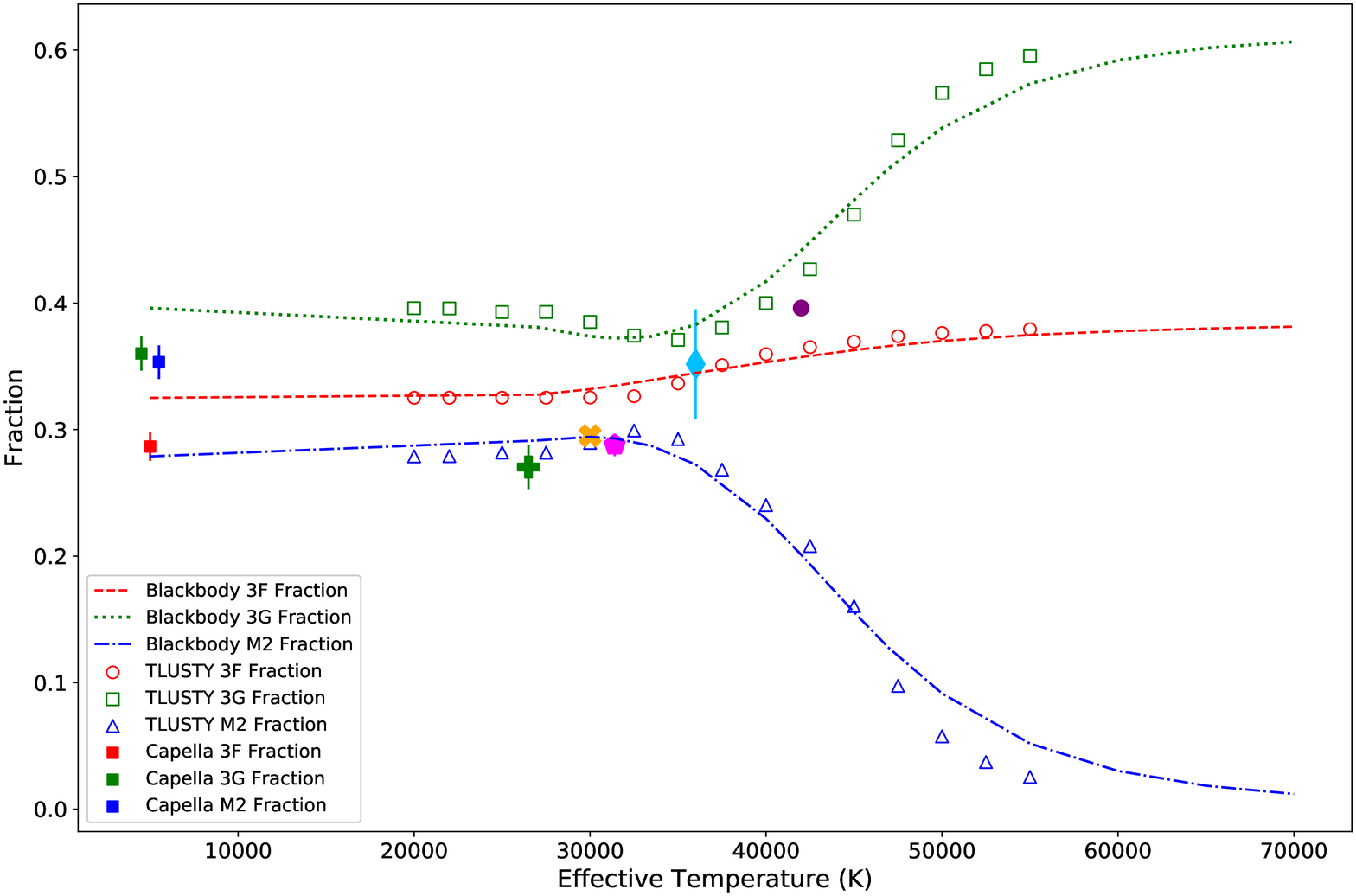}
\caption{\ion{Fe}{17} fractional line strengths as a function of UV field temperature. The dashed lines represent the FAC-calculated fractions; the open points represent the fractions calculated using TLUSTY model atmospheres (combination of BSTAR2006 and OSTAR2002 data sets); the leftmost points represent the observed 3F, 3G, and M2 fractions of Capella ($T_{\text{eff}}$ = 5000 K), and the larger symbols represent (left to right) the observed 3F fractions of $\epsilon$~Ori ($T_{\text{eff}}$ = 27000 K), $\zeta$~Ori ($T_{\text{eff}}$ = 30000 K), $\tau$~Sco ($T_{\text{eff}}$ = 31400 K), $\lambda$~Cep ($T_{\text{eff}}$ = 36000 K), and $\zeta$~Pup ($T_{\text{eff}}$ = 42500 K).
\label{fig:zetacomp}}
\end{figure*}

We also estimated the line strengths using Gaussian fits for comparison to predicted ratios in Figures~\ref{fig:crm_3F} and \ref{fig:zetacomp}. Figure~\ref{fig:zpgauss} shows our fits to the XMM-Newton RGS spectra of $\zeta$~Pup using Gaussian models.

For comparison to stars of similar spectral type, we also fit Gaussians to the spectra of the OB supergiants $\epsilon$~Ori, $\zeta$~Ori, and $\lambda$~Cep.  \cite{Rauw2015} previously found a similar but weaker anomaly in the $2p\mbox{ -- }3s$ line ratios of $\lambda$~Cep as in $\zeta$~Pup. $\zeta$~Ori and $\epsilon$~Ori are statistically consistent with Capella, and our model predicts only a slight deviation with respect to Capella. $\lambda$~Cep does have a somewhat stronger best-fit 3F fraction, but as it is also much fainter due to its larger distance, the statistical uncertainties on the fractions are large, and the spectrum is marginally consistent with the 3F fractions observed in both Capella and $\zeta$~Pup.

The uncertainties shown in Figures~\ref{fig:crm_3F} and \ref{fig:zetacomp} are statistical only. A few systematic effects are possible, and we consider those here. 

First, the weak continuum flux, mainly due to bremsstrahlung, is estimated by fitting nearby spectral regions that are free of lines. We tried changing the estimated continuum flux by $\pm 25$\% and found systematic effects no larger than the statistical uncertainties. 

Second, weak, blended lines might contaminate the ratio measurements. The most likely such lines would be from the Rydberg series of helium-like oxygen. We did not make a quantitative estimate of this effect, although given the strengths of the unblended lower-$n$ lines in the series, it cannot be too important. We note that this should affect all of the massive stars about equally, so the difference in 3F fraction observed between $\zeta$~Pup and the other stars is still a robust conclusion.

Third, the line shapes of massive stars are not Gaussian but have a skewed shape resulting from differential absorption of distributed X-ray emission in their supersonic winds \citep{OC01,Cel06}. Nevertheless, the error in {\it total line flux} resulting from fitting a Gaussian to such lines is actually quite small, as can be seen in the residuals in Figure~\ref{fig:zpgauss}.

Finally, we have not accounted for differential wind absorption effects due to slightly different photoelectric continuum opacities at the respective wavelengths of 3F, 3G, and M2 \citep{windtabs2010}. Stars with higher wind optical depths at these wavelengths, such as $\zeta$~Pup, can be treated approximately in the exospheric limit, where the emergent flux at a given wavelength scales approximately inversely with the opacity. This scaling holds for arbitrarily high optical depth and thus gives an upper limit to the correction to the observed line ratios. We estimate this effect at approximately 5\% between 3F versus 3G and M2. While this is a significant correction that partially mitigates the observed discrepancy, we stress that, by itself, it cannot solve the issue.

In Figure~\ref{fig:zetacomp}, we compare the fractions for $\epsilon$~Ori, $\zeta$~Ori, $\lambda$~Cep, and $\zeta$~Pup to the FAC-predicted fractions at maximum UV photoexcitation ($W = 0.5$) as a function of blackbody temperature. The fractional values for Capella are shown in Figure~\ref{fig:zetacomp} to represent points in the limit of no photoexcitation. We examined the M2~/~(3F~+~3G~+~M2) fraction as a diagnostic of the quality of our line intensity calculations. We observed a $\sim$20$\%$ model-data discrepancy for the M2 fraction, which is consistent with the discrepancies found in similar previous studies (see \S~\ref{sec:fexvii}).

However, the FAC-predicted 3F fraction calculation does not accurately reproduce the observed ratios. As stated in \S~\ref{sec:atomic_model}, the FAC-predicted dynamic range for the 3F fraction (as a function of temperature) is not large enough to explain the observed ratio difference between Capella and $\zeta$~Pup.

\section{Discussion}
\label{sec:disc}

The discrepancy between the measured Capella and $\tau$ Sco M2 fractions and our FAC CRM fraction calculations is approximately 20$\%$, which is consistent with several previous studies that examined these line ratios \citep{Loch06, Gu2008}.  However, the discrepancy between the 3F fractions of Capella and $\zeta$~Pup is both significant and surprising. 

We were unable to reproduce the observed difference between the ratios of the two stars in the FAC-predicted dynamic range of the ratios as a function of UV field intensity. As illustrated in Figure~\ref{fig:zetacomp}, the dynamic range of the calculated 3F fraction is less than what we observed between Capella and $\zeta$~Pup. 

We considered several other possible atomic and astrophysical processes in attempts to potentially explain the discrepancy in the 3F~/~(3F~+~3G~+~M2) fraction in $\zeta$~Pup. 

We considered possible contamination from fluorine lines, as the wavelengths of the strongest hydrogen-like and helium-like fluorine lines are very close to the wavelengths of the $2p-3s$ and $2p-3d$ \ion{Fe}{17} transitions \citep{Beiersdorfer2017ApJ}. Fluorine typically has a very low abundance in most astrophysical objects, and its lines are thus usually negligible in X-ray spectra. But as $\zeta$~Pup is known to show strong surface enhancement of nitrogen from CNO processed material \citep[e.g.,][]{Bouret2012}, one might speculate that fluorine could also be enhanced through higher-temperature analogs of the CNO cycle. We tried including the lines of helium-like \ion{F}{8} using the {\tt hewind} model in conjunction with the $2p-3s$ lines of \ion{Fe}{17} in our model fitting, but we found that the model fit strongly preferred to have zero intensity for fluorine lines, and forcing the fluorine line flux to be nonnegligible produced clearly unacceptable model fits. We thus conclude that the fluorine abundance is indeed negligible, and fluorine lines cannot explain the anomalous 3F line strength in $\zeta$~Pup.

We also considered resonant Auger destruction \citep{Liedahl2005} of the 3G and/or M2 lines by $2p-3d$ transitions in the low charge states of Fe dominant in the bulk of the wind, likely \ion{Fe}{4} -\ion{}{6}, as proposed in \cite{ML2012}. The $2p-3d$ transition energies for these charge states are in the range 17.0-17.2 $\mbox{\AA}$, but are not sufficiently well known to evaluate the degree of coincidence with 3G and M2 \citep{Gu2006,Blancard_2018}, so it is not possible to confirm or rule out this possibility. For this effect to explain the observed ratio of \ion{Fe}{17} $2p-3s$ lines in $\zeta$~Pup while having a negligible effect for other OB stars, the relevant transition would have to be in a charge state of iron that is more prevalent in the wind of $\zeta$~Pup than for other stars.

\cite{BSO2003} showed that strong magnetic fields can induce direct decay of the $J=0$ 3$s$ excited state to ground; the transition wavelength of 16.804~\AA\ is close enough to the wavelength of 3F that for the velocity broadened winds of OB stars, the flux of this transition would blend with 3F. Thus, this could effectively remove photons from 3G and appear to feed 3F. This is unlikely to be relevant for $\zeta$~Pup, since the required magnetic field to produce an effect is $\sim$50 kG, while observational limits to the photospheric magnetic field strength of $\zeta$~Pup are at the level of $\sim$30 G for a global dipole configuration \citep{David-Uraz2014}, or $\sim$ kG for small-scale disordered fields \citep{Kochukov2013}. Furthermore, as we showed in \S~\ref{sec:atomic_model}, the rate of feeding of the $J=0$ $3s$ excited state is not sufficient to explain the enhanced strength of 3F.

We showed in \S~\ref{sec:atomic_model} that strongly adjusting the relative rate of population of the $J=0$ 3$s$ excited state could produce a sufficient dynamic range in the strength of 3F to explain the observed ratio in $\zeta$~Pup. There is no reason to think that FAC should strongly underestimate this rate, but one might suspect that perhaps inadequate treatment of configuration mixing might cause such an issue.
We therefore tested the impact of including mixing between the ground state and singly excited states with $n$ = {3, 4} and $n$ = {3, 4, 5}, such that the calculation includes the levels arising from the $2s^{2} 2p^{6}$, $2s2p^{6}nl$, and $2s^{2} 2p^{5}nl$ configurations. The calculations including more mixing had only a very small effect on the dynamic range of 3F and the population of the $J=0$ $3s$ excited state. Although configuration mixing is known to converge rather slowly, the most important effects should occur when including the levels we studied, so the absence of a significant improvement in agreement with observations indicates that this is not the origin of the discrepancy.

\cite{Pollock2007} has suggested that charge exchange (CX) could be important in O-star winds. One might suppose that this could alter the line ratios of \ion{Fe}{17} in a way that might reproduce the observations, although it would not be clear why this should preferentially affect $\zeta$~Pup over other O stars. \cite{Betancourt2018} have measured X-ray spectra of neon-like Ni, which can be taken as a crude proxy for the expected spectrum of neon-like Fe. Their measurements show that the M2 line is dominant and that this is a simple consequence of cascade probabilities for most of the highly excited states populated by CX. If CX were important in the wind of $\zeta$~Pup, this would thus only enhance the discrepancy between modeled and observed $2p-3s$ line ratios. We thus conclude that CX is likely negligible, and in any case cannot explain the anomalous strength of 3F in $\zeta$~Pup.

Considering the thorough measures incorporated by previous studies (see \S~\ref{sec:fexvii}), we believe that using a more comprehensive model would not significantly change the dynamic range of 3F~/~(3F~+~3G~+~M2). A better model would likely enhance feeding of 3G and M2 (for all UV fluxes) and thus produce better agreement for, e.g. Capella, as shown in \cite{Gu2008}. However, these improvements would not completely mitigate the model-data discrepancy.

In terms of future directions, laboratory astrophysics experiments could be the key to solving the model-data discrepancy. 
A potential future study could improve the FLASH-EBIT measurements~\citep{Shah19} to derive better constraints for the \ion{Fe}{17} 3$s$ cross section by using, for example, a high-resolution wide-band X-ray microcalorimeter. 
Furthermore, a potential future laboratory experiment could use simultaneous EUV and X-ray spectroscopy to benchmark the importance of the 3$s$ $J$ = 0 level in feeding the upper level of the 3G line, as well as the relative strengths of the $3s\mbox{ -- }3p$ transitions. If the modeled rate of the feeding of the 3$s$ $J$ = 0 level were strongly underestimated, this may potentially reconcile the discrepancy.

\section{Conclusions}
We used the FAC collisional-radiative model to model the effect of UV photoexcitation from O stars on the \ion{Fe}{17} $2p\mbox{ -- }3s$ line ratios. We solved the rate equations, deriving an analytical model to calculate the ratios as a function of UV field intensity using parameters derived from the FAC CRM, experiments, and astrophysical observations. Using these models, we demonstrated that the UV field intensities of O stars have at most a marginal effect on the line ratios. We also implemented a line profile model for \ion{Fe}{17} in the spectra of O stars, called {\tt newind}, in analogy with the {\tt hewind} model of \cite{LPKC06}, where the profiles are calculated including the radial dependence of the $2p\mbox{--}3s$ ratios.

We compared our model calculations to archival observations of coronal and hot stars taken by Chandra and XMM-Newton. The comparison with Capella showed model-data discrepancies consistent with ones found in previous studies. More importantly, the dynamic range of the model 3F fraction as a function of UV field intensity was not large enough to explain the observed difference in this ratio between Capella and $\zeta$~Pup. We conclude that UV photoexcitation has only a weak effect on the line ratios of \ion{Fe}{17} in O stars and that it cannot explain the observed strength of 3F in $\zeta$~Pup. Future laboratory experiments could potentially use simultaneous EUV and X-ray spectroscopy both to place constraints on the \ion{Fe}{17} 3$s$ cross sections and to benchmark the importance of the 3$s$ $J$ = 0 level in feeding the upper level of the 3G line by measuring the strength of the 1153 \AA\ line relative to the strengths of the $2p\mbox{--}3s$ lines.

\acknowledgements

We thank Ming Feng Gu for stimulating discussions. We thank Ehud Behar and an anonymous referee for their helpful comments. We acknowledge support from NASA's Astrophysics division. G.G. acknowledges support under NASA award No. 80GSFC21M0002. C.S. acknowledges the support by an appointment to the NASA Postdoctoral Program at the NASA Goddard Space Flight Center, administered by Universities Space Research Association under contract with NASA, by the Lawrence Livermore National Laboratory (LLNL) Visiting Scientist and Professional Program Agreement No.s\ VA007036 and VA007589, and by Max-Planck-Gesellschaft (MPG). 

\bibliographystyle{aasjournal}
\bibliography{Master,RIES}

\begin{appendices}
\appendix
\vspace{-8mm}
\section{Archival Observations}
\label{appendix:obsid}

\begin{table}[ht!]
\centering
\setlength{\tabcolsep}{12pt}
  \caption{Log of archival observation data. OBSIDs with a * sign denote spectra obtained from the {\it Chandra} archive, and those with a + sign denote spectra obtained from the {\it XMM-Newton} Science Archive.
\label{tab:obsid_appendix}}
    \begin{tabular}{ccc}
    \hline\hline
      Star & OBsID & Exposure (ks)\\ 
      \hline
      Capella & $1099^{*}$ & 14.6\\
      Capella & $3674^{*}$ & 28.7\\
      Capella & $6471^{*}$ & 29.6\\
      \hline
      $\epsilon$ Orionis & $3753^{*}$ & 91.7\\
      $\epsilon$ Orionis & $0112400101^{+}$ & 12.9\\
      \hline
      $\zeta$ Orionis & $610^{*}$ & 59.7\\
      $\zeta$ Orionis & $13460^{*}$ & 142.9\\
      $\zeta$ Orionis & $0112530101^{+}$ & 42.0\\
      $\zeta$ Orionis & $0657200101^{+}$ & 97.8\\
      $\zeta$ Orionis & $0657200201^{+}$ & 47.4\\
      $\zeta$ Orionis & $0657200301^{+}$ & 43.9\\
      \hline
      EX Hydra & $1706^{*}$ & 150.6\\
      \hline
      $\tau$ Scorpii & $638^{*}$ & 59.2\\
      $\tau$ Scorpii & $2305^{*}$ & 13.0\\
      $\tau$ Scorpii & $0112540101^{+}$ & 23.2\\
      $\tau$ Scorpii & $0112540201^{+}$ & 10.8\\
      \hline
      $\lambda$ Cephei & $0720090301^{+}$ & 75.8\\
      $\lambda$ Cephei & $0720090401^{+}$ & 82.4\\
      $\lambda$ Cephei & $0720090501^{+}$ & 94.7\\
      $\lambda$ Cephei & $0720090601^{+}$ & 15.7\\
      \hline
      $\zeta$ Puppis & $0095810301^{+}$ & 52.5\\
      $\zeta$ Puppis & $0095810401^{+}$ & 39.9\\
      $\zeta$ Puppis & $0157160401^{+}$ & 41.5\\
      $\zeta$ Puppis & $0095810501^{+}$ & 38.7\\
      $\zeta$ Puppis & $0095810901^{+}$ & 43.5\\
      $\zeta$ Puppis & $0157161101^{+}$ & 27.8\\
      $\zeta$ Puppis & $0159360101^{+}$ & 66.2\\
      $\zeta$ Puppis & $0159360301^{+}$ & 26.9\\
      $\zeta$ Puppis & $0159360501^{+}$ & 34.5\\
      $\zeta$ Puppis & $0159360701^{+}$ & 23.5\\
      $\zeta$ Puppis & $0159360901^{+}$ & 48.3\\
      $\zeta$ Puppis & $0159361101^{+}$ & 42.5\\
      $\zeta$ Puppis & $0159361301^{+}$ & 61.1\\
      $\zeta$ Puppis & $0163360201^{+}$ & 41.6\\
      $\zeta$ Puppis & $0414400101^{+}$ & 58.3\\
      $\zeta$ Puppis & $0561380101^{+}$ & 64.1\\
      $\zeta$ Puppis & $0561380201^{+}$ & 76.6\\
      $\zeta$ Puppis & $0561380301^{+}$ & 63.7\\
      $\zeta$ Puppis & $0561380501^{+}$ & 60.5\\
      $\zeta$ Puppis & $0561380601^{+}$ & 67.5\\
      $\zeta$ Puppis & $0561380701^{+}$ & 55.0\\
      \hline\hline
    \end{tabular}
\end{table}

\clearpage

\section{Solution of rate equations}
\label{appendix:rate}

In this appendix we solve the rate equations to obtain line ratios for the $3s-2p$ transitions of \ion{Fe}{17} as a function of UV field strength and density. The levels are labeled using the level numbers from FAC (as shown in Table \ref{tab:lvls}) in ascending energy order, so that the ground state is 0, the four $2p^{-1} 3s$ excited states are 1-4 (called $3s$ in this appendix for short), the 10 $2p^{-1} 3p$ excited states are 5-14 (called $3p$), the 12 $2p^{-1} 3d$ excited states are 15-26 (called $3d$), and the two $2s^{-1} 3s$ states are 27 and 28 (called $3s^\prime$; although note that level 28 has only very small oscillator and collision strengths connecting it to levels 1 and 3, so its participation is negligible). First, we give the equation for level 1 (upper level of the M2 line):

\begin{equation}
\begin{split}
\frac{dN_1}{dt} &= n_e N_0 C_1 - N_1 A_{1,0} + \sum_{i=3p,3s^\prime} (N_i A_{i,1} - N_1 \phi_{1,i}) - \sum_{j=3s,3p,3d,3s^\prime}^{j \ne 1} n_e N_1 C_{1,j} + n_e N_3 C_{3,1} \\ 
&= R_1 - N_1 A_1 + \sum_{i=3p,3s^\prime} (N_i A_{i,1} - N_1 \phi_{1,i}) - \sum_{j=3s,3p,3d,3s^\prime}^{j \ne 1} n_e N_1 C_{1,j} + n_e N_3 C_{3,1}\, .
\label{eq:level1}
\end{split}
\end{equation}

Here we use the following notation: $n_e$ is the electron density (cm$^{-3}$); $N_i$ is the density of ions in level $i$; $C_{i,j}$ (cm$^3$ s$^{-1}$) is the collisional rate coefficient from level $i$ to level $j$; $C_i$ (cm$^3$ s$^{-1}$) is the collisional rate coefficient from the ground state to excited state $i$ , but using the shorthand that it includes {\it all} processes populating that level (i.e. cascades, etc.), other than those from the $3s$, $3p$, $3d$, and $3s^\prime$ levels, which are all explicitly accounted for; $R_i$ (cm$^{-3}$ s$^{-1}$) is analogous to $C_i$, but includes the electron and ion ground-state densities, i.e., $R_i \equiv n_e N_0 C_i$; $A_{i,j}$ (s$^{-1}$) is the spontaneous decay rate from level $i$ to level $j$, and $A_i$ is the total decay rate from level $i$ summed over all $j$; and $\phi_{i,j}$ (s$^{-1}$) is the photoexcitation rate from level $i$ to level $j$. 


We include electron collisions within $n=3$ excited states, including excitations from $3s$ states to $3s^\prime$ states (i.e. $2s-2p$ excitations), but we neglect collisions to $n=4$ and higher. We also neglect photoexcitation between $3s$ states, since the oscillator strengths are very small. We neglect photoexcitation to $4p$ states, since the energies are much larger.


Thus, from left to right, the terms in Equation~\ref{eq:level1} represent all direct collisions, as well as cascades that are not explicitly accounted for; decays to ground; decays from and photoexcitations to $3p$ and $3s^\prime$; collisional excitation to $3s$, $3p$, $3d$, and $3s^\prime$; and collisional de-excitation from level 3. 

Similarly, these are the rate equations for the other $3s$ states, levels 2-4:

\begin{equation}
\frac{dN_2}{dt} = R_2 + N_3 A_3 - N_2 A_2 + \sum_{i=3p,3s^\prime} (N_i A_{i,2}) + n_e N_1 C_{1,2} + n_e N_3 C_{3,2}\, ;
\end{equation}

\begin{equation}
\frac{dN_3}{dt} = R_3 - N_3 A_3 + \sum_{i=3p,3s^\prime} (N_i A_{i,3} - N_3 \phi_{3,i}) - \sum_{j=3s,3p,3d,3s^\prime}^{j \ne 3} n_e N_3 C_{3,j} + n_e N_1 C_{1,3}\, ;
\end{equation}

\begin{equation}
\frac{dN_4}{dt} = R_4 - N_4 A_4 + \sum_{i=3p,3s^\prime} (N_i A_{i,4}) + n_e N_1 C_{1,4} + n_e N_3 C_{3,4}\, .
\end{equation}

For each $3s$ level, the total $A_i$ is dominated by decay to a single level: for level 3, the branching ratio to level 1 is just $\sim 10^{-4}$, and decay to ground is strictly forbidden, so level 2 is the dominant decay channel (leading to the term $N_3 A_3$ in the equation for level 2); while for levels 1, 2, and 4, the ground state is the dominant decay channel. Photoexcitation and collisional excitation from levels 2 and 4 to the $3p$ manifold are neglected, since these states are not metastable.

These are the rate equations for the $3p$ states, levels 5-14:
\begin{equation}
\frac{dN_i}{dt} = R_i - N_i A_i + \sum_{j=1,3} N_j (\phi_{j,i} + n_e C_{j,i}) + \sum_{k=3d} N_k A_{k,i}\, .
\end{equation}
For levels 6, 9, and 13, direct decay to the ground state is nonnegligible, although the branching fraction is still small. The final term accounts for $3d-3p$ decays. 

Similarly, for $3d$ states, levels 15-26:
\begin{equation}
\frac{dN_i}{dt} = R_i - N_i A_i + \sum_{j=1,3} N_j (n_e C_{j,i})\, ,
\end{equation}
and for the $3s^\prime$ states, levels 27-28,
\begin{equation}
\frac{dN_i}{dt} = R_i - N_i A_i + \sum_{j=1,3} N_j (\phi_{j,i} + n_e C_{j,i})\, ,
\end{equation}

The $3d$ states that are relevant primarily radiatively decay to $3p$ states, although level 16 has a nonnegligible branching fraction to ground, but the $3s^\prime$ states only radiatively decay to $3s$ states. 

Since we assume a steady state, all of the derivatives on the left-hand side of the equations equal zero. Thus, we can solve for the population of $3d$ level $i$:

\begin{equation}
N_i = \frac{R_i + \sum_{j=1,3} N_j (n_e C_{j,i})}{A_i}\, ,
\end{equation}
and $3s^\prime$ level $i$:
\begin{equation}
N_i = \frac{R_i + \sum_{j=1,3} N_j (\phi_{j,i}+ n_e C_{j,i})}{A_i}\, .
\end{equation}

We can similarly solve for the $3p$ levels, substituting the expressions for $3d$ level populations
\begin{equation}
N_i = \frac{R_i + \sum_{j=1,3} [N_j (\phi_{j,i} + n_e C_{j,i})] + \sum_{k=3d} [R_k + \sum_{j=1,3} N_j (n_e C_{j,k})] F_{k,i}}{A_i}\, ,
\end{equation}
where $F_{i,j} = A_{i,j} / A_{i}$ is the branching ratio to lower level $j$ from upper level $i$ (in this case, from $3d$ to $3p$ levels). 

We can then substitute the expressions for $N_i$ of the $3p$ and $3s^\prime$ levels in the equation for $N_3$, again setting the derivative to zero:
\begin{equation}
\begin{split}
N_3 A_3 &= R_3 + \sum_{i=3p} \big[\big(R_i + \sum_{j=1,3} \{N_j (\phi_{j,i} + n_e C_{j,i})\} + \sum_{k=3d} [R_k + \sum_{j=1,3} N_j (n_e C_{j,k})] F_{k,i}\big)F_{i,3} - N_3 \phi_{3,i}\big] \\
&+ \sum_{i=3s^\prime} \big[\big(R_i + \sum_{j=1,3} N_j (\phi_{j,i} + n_e C_{j,i})\big) F_{i,3} - N_3 \phi_{3,i}\big]\, - \sum_{j=3s,3p,3d,3s^\prime}^{j \ne 3} n_e N_3 C_{3,j} + n_e N_1 C_{1,3}\, .
\end{split}
\end{equation}

Define $R^\prime_j = R_j + \sum_{i=3p} [R_i F_{i,j} + \sum_{k=3d} R_k F_{k,i} F_{i,j}] + \sum_{i=3s^\prime} R_i F_{i,j}$; this adds to $R_j$ the line strength from the $3p$ channel that is expected to go to level $j$ in the absence of photoexcitation, including cascades through $3p$ originating in $3d$, as well as cascades to $3s$ from $3s^\prime$ states. Then,

\begin{equation}
    \begin{split}
        N_3 A_3 &= R^\prime_3 + \sum_{i=3p} \big[\sum_{j=1,3} \{N_j (\phi_{j,i} + n_e C_{j,i} + \sum_{k=3d} n_e C_{j,k} F_{k,i}) F_{i,3}\} - N_3 \phi_{3,i}\big] \\
&+ \sum_{i=3s^\prime} \sum_{j=1,3} \big[N_j (\phi_{j,i} + n_e C_{j,i}) F_{i,3} - N_3 \phi_{3,i}\big] - \sum_{j=3s,3p,3d,3s^\prime}^{j \ne 3} n_e N_3 C_{3,j} + n_e N_1 C_{1,3} \\ 
&= R^\prime_3 + \sum_{i=3p} \big[N_1 \big(\phi_{1,i} + n_e (C_{1,i} + \sum_{k=3d} C_{1,k} F_{k,i})\big)F_{i,3} - N_3 \big(\phi_{3,i}(1-F_{i,3}) + n_e \{C_{3,i}(1-F_{i,3}) \\
&+ \sum_{k=3d} C_{3,k} (1 - F_{k,i} F_{i,3})\}\big) \big] \\
&+ \sum_{i=3s^\prime} \big[N_1 (\phi_{1,i} + n_e C_{1,i}) F_{i,3} - N_3 (\phi_{3,i} + n_e C_{3,i}) (1 - F_{i,3})\big] - \sum_{j=3s}^{j \ne 3} n_e N_3 C_{3,j} + n_e N_1 C_{1,3}\, .
    \end{split}
\end{equation}

Moving terms with $N_3$ to the left, we have
\begin{equation}
    \begin{split}
        &N_3 [A_3 + \sum_{i=3p} \{\phi_{3,i} (1 - F_{i,3}) + n_e(C_{3,i} (1 - F_{i,3}) + \sum_{k=3d} C_{3,k} (1 - F_{k,i} F_{i,3}))\} \\
&+ \sum_{i=3s^\prime} (\phi_{3,i} + n_e C_{3,i}) (1 - F_{i,3}) + \sum_{j=3s}^{j \ne 3} n_e C_{3,j}] \\
&= R^\prime_3 + N_1 \big[\sum_{i=3p} \big(\phi_{1,i} + n_e (C_{1,i} + \sum_{k=3d} C_{1,k} F_{k,i})\big) F_{i,3} + \sum_{i=3s^\prime} (\phi_{1,i} + n_e C_{1,i}) F_{i,3} + n_e C_{1,3}\big]\, .
    \end{split}
\end{equation}

Now define 
\begin{equation}
    P_{i,j} \equiv \frac{1}{A_i} \sum_{k=3p,3s^\prime} \phi_{i,k} F_{k,j}\, ,
\end{equation}
and 
\begin{equation}
    P_i \equiv \frac{1}{A_i} \sum_{k=3p,3s^\prime} \phi_{i,k} (1 - F_{k,i})\, .
\end{equation}
$P_{i,j}$ gives the effective photoexcitation rate from $3s$ level $i$ to $3s$ level $j$, summed over all intermediate $3p$ and $3s^\prime$ states, and normalized to the decay rate from level $i$, $A_i$; while $P_i$ similarly gives the effective normalized photoexcitation rate from $3s$ level $i$ to {\it all other} levels combined. For the latter, this also includes the small fraction of direct decays to ground from $3p$ levels. Photoexcitation via $3s^\prime$ levels can usually be neglected, since the transition energies to these levels are much higher than to $3p$ levels, and the UV flux is thus much lower. 

Similarly, define
\begin{equation}
    \frac{1}{n_{i,j}} \equiv \frac{1}{A_i} \big[\sum_{k=3p} \big(C_{i,k} + \sum_{m=3d} C_{i,m} F_{m,k}\big) F_{k,j} + \sum_{k=3s^\prime}  C_{i,k} F_{k,j} + C_{i,j}\big]\, ,
\end{equation}
and
\begin{equation}
    \frac{1}{n_{i}} \equiv \frac{1}{A_i} \big[\sum_{k=3p} \big(C_{i,k} (1 - F_{k,i}) + \sum_{m=3d} C_{i,m} (1 - F_{m,k} F_{k,i})\big) + \sum_{k=3s^\prime} C_{i,k} (1 - F_{k,i}) + \sum_{k=3s}^{k \ne i} C_{i,k}\big] \, .
\end{equation}
Parameters $n_{i,j}$ and $n_i$ are effectively critical densities, with $n_e / n_{i,j}$ giving the effective rate of collisional feeding from level $i$ to level $j$ summed over all intermediate states, and normalized to $A_i$; and with $n_e / n_i$ summing over all states other than the initial state $i$.

Using the definitions for $P_{i,j}$, $P_i$, $n_{i,j}$, and $n_i$, we can write the expressions for $N_3$ and $N_1$ as 
\begin{equation}
N_3 A_3 = \frac{R^\prime_3 + N_1 A_1 [P_{1,3} + n_e / n_{1,3}]}{1 + P_3 + n_e / n_3}\, ;
\end{equation}
\begin{equation}
N_1 A_1 = \frac{R^\prime_1 + N_3 A_3 [P_{3,1} + n_e / n_{3,1}]}{1 + P_1 + n_e / n_1}\, .
\end{equation}

Then, solve the two equations to obtain independent expressions for $N_3$ and $N_1$:
\begin{equation}
N_3 A_3 = \frac{R^\prime_3 (1 + P_1 + n_e / n_1) + R^\prime_1 [P_{1,3} + n_e / n_{1,3}]}{(1 + P_1 + n_e / n_1) (1 + P_3 + n_e / n_3) - [P_{3,1} + n_e / n_{3,1}] [P_{1,3} + n_e / n_{1,3}]}\, ;
\end{equation}
\begin{equation}
N_1 A_1 = \frac{R^\prime_1 (1 + P_3 + n_e / n_3) + R^\prime_3 [P_{3,1} + n_e / n_{3,1}]}{(1 + P_1 + n_e / n_1) (1 + P_3 + n_e / n_3) - [P_{3,1} + n_e / n_{3,1}] [P_{1,3} + n_e / n_{1,3}]}\, .
\end{equation}

One can obtain similar expressions for levels 2 and 4, which are left in terms of the populations of levels 1 and 3:
\begin{equation}
N_2 A_2 = R^\prime_2 + N_1 A_1 (P_{1,2} + n_e / n_{1,2}) + N_3 A_3 (1 + P_{3,2} + n_e / n_{3,2})\, ;
\end{equation}
\begin{equation}
    N_4 A_4 = R^\prime_4 + N_1 A_1 (P_{1,4} + n_e / n_{1,4}) + N_3 A_3 (P_{3,4} + n_e / n_{3,4}) \, .
\end{equation}

Next, consider the line strengths in the limit of zero photoexcitation and low density, i.e. all $P_i$ and $P_{i,j}$ are zero, and $n_e \ll n_i$ and $n_{i,j}$. Then, $N_1 A_1 = R^\prime_1$; $N_2 A_2 = R^\prime_2 + R^\prime_3$; $N_3 A_3 = R^\prime_3$; and $N_4 A_4 = R^\prime_4$. It is not an accident or overcount that $R^\prime_3$ shows up in the expressions for levels 2 and 3; this is because electrons populating level 3 will decay twice, first from level 3 to 2, emitting a 10.8 eV photon, and then from level 2 to 0, emitting a 727.1 eV photon. 

The observable line ratios for levels 1, 2, and 4 can be calculated as
\begin{equation}
\mathcal{R}_i = \frac{N_i A_i}{\sum_{j=1,2,4} N_j A_j}\, .
\label{eq:ratio_def}
\end{equation}
One can also evaluate this expression for $\mathcal{R}_3$; it gives the ratio of the flux in the 10.8 eV line to the sum of the three $3s-2p$ lines.

In the limit of low density and UV flux, the denominator sums to $\sum^4_{i=1} R^\prime_i$.
Define $\mathcal{R}^\circ_i = R^\prime_i / \sum^4_{j=1} R^\prime_j$; these quantities give the values of $\mathcal{R}_i$ in the limit of low density and low UV flux, with the exception that $\mathcal{R}_2 = \mathcal{R}^\circ_2 + \mathcal{R}^\circ_3$.

Provisionally neglect the weak decay channels from levels 6, 9, 13, and 16 to ground; in this case the denominator of Equation~\ref{eq:ratio_def} is a constant. Then,

\begin{equation}
        \mathcal{R}_1 = \frac{\mathcal{R}^\circ_1 (1+ P_3 + n_e / n_3) + \mathcal{R}^\circ_3 (P_{3,1} + n_e / n_{3,1})}{(1 + P_1 + n_e / n_1) (1 + P_3 + n_e / n_3) - (P_{3,1} + n_e / n_{3,1}) (P_{1,3} + n_e / n_{1,3})} \\
\end{equation}
\begin{equation}
\mathcal{R}_2 = \mathcal{R}^\circ_2 + \mathcal{R}_1 (P_{1,2} + n_e / n_{1,2}) + \mathcal{R}_3 (1 + P_{3,2} + n_e / n_{3,2})\, ; 
\end{equation}
\begin{equation}
    \mathcal{R}_3 = \frac{\mathcal{R}^\circ_3 (1+ P_1 + n_e / n_1) + \mathcal{R}^\circ_1 (P_{1,3} + n_e / n_{1,3})}{(1 + P_1 + n_e / n_1) (1 + P_3 + n_e / n_3) - (P_{3,1} + n_e / n_{3,1}) (P_{1,3} + n_e / n_{1,3})}\, ; 
\end{equation}
\begin{equation}
    \mathcal{R}_4 = \mathcal{R}^\circ_4 + \mathcal{R}_1 (P_{1,4} + n_e / n_{1,4}) + \mathcal{R}_3 (P_{3,4} + n_e / n_{3,4}) \, .
\end{equation}
Again, the expressions for levels 2 and 4 can be evaluated using the results for levels 1 and 3. 

Now let us consider the effect of neglecting direct decays to ground of $3p$ and $3d$ excited states in these expressions. Since the denominator in Equation~\ref{eq:ratio_def} is the total $3s$ line strength in the limit of low density and UV flux, in general the resulting line ratios will sum to slightly less than unity, with the remainder accounted for by strengthening of the $2p-3p$ and $2p-3d$ transitions. Since we are only interested in the {\it relative} strengths of the $2p-3s$ transitions, we can simply renormalize the ratios to sum to unity.

The values for $\mathcal{R}^\circ_i$ can be fixed from theory, evaluated at low UV flux and low density; or from experiments at sufficiently low density; or from astrophysical observations of low-density, low-UV flux objects. The values for $P_{i,j}$ and $P_i$ can be calculated given the relevant oscillator strengths, $A$ values, and branching ratios, as well as the input UV spectrum, and similarly for $n_{i,j}$ and $n_i$ with collision strengths, which are weakly temperature dependent. Finally, assuming trivial radiative transfer in the circumstellar environment, a common geometrical dilution term $W(r)$ can be factored out from all $P$; or if not, $J_\nu (r)$ must be evaluated for all relevant wavelengths over the circumstellar environment.

\end{appendices}
\end{document}